\documentclass{aastex631}

\usepackage{CJK}
\usepackage{listings}

\newcommand{\teff}{T_{\rm eff}}
\newcommand{\logg}{\log{g}}
\newcommand{\meta}{{\rm [M/H]}}
\newcommand{\Rv}{R_{\rm V}}

\newcommand{\BPRP}{G_{\rm BP}\,{-}\,G_{\rm RP}}
\newcommand{\BPK}{G_{\rm BP}\,{-}\,K_{\rm S}}
\newcommand{\JK}{J\,{-}\,K_{\rm S}}
\newcommand{\EBPRP}{E(G_{\rm BP}\,{-}\,G_{\rm RP})}
\newcommand{\EBPK}{E(G_{\rm BP}\,{-}\,K_{\rm S})}
\newcommand{\EJK}{E(J\,{-}\,K_{\rm S})}
\newcommand{\BPRPO}{(G_{\rm BP}\,{-}\,G_{\rm RP})_0}
\newcommand{\BPKO}{(G_{\rm BP}\,{-}\,K_{\rm S})_0}
\newcommand{\JKO}{(J\,{-}\,K_{\rm S})_0}
\newcommand{\gspphot}{GSP-phot}
\newcommand{\Ks}{K_{\rm S}}
\newcommand{\EBV}{E(B{-}V)}

\defcitealias{Andrae2023}{A23}
\defcitealias{ZhangXY2023}{Z23}
\defcitealias{Green2019}{G19}

\def\mathbi#1{\textbf{\em #1}}

\begin{document}
\begin{CJK*}{UTF8}{gbsn}

\title{Data-driven stellar intrinsic colors and dust reddenings for spectro-photometric data: \\
From the blue-edge method to a machine-learning approach}

\correspondingauthor{He Zhao \& Shu Wang}
\email{hzhao@pmo.ac.cn, shuwang@nao.cas.cn}

\author[0000-0003-2645-6869]{He Zhao (赵赫)}
\affiliation{Purple Mountain Observatory and Key Laboratory of Radio Astronomy, Chinese Academy of Sciences, 10 Yuanhua Road, Nanjing 210033, People's Republic of China}

\author[0000-0003-4489-9794]{Shu Wang (王舒)}
\affiliation{CAS Key Laboratory of Optical Astronomy, National Astronomical Observatories, Chinese Academy of Sciences, Beijing 100101, People's Republic of China}

\author[0000-0003-3168-2617]{Biwei Jiang (姜碧沩)}
\affiliation{Institute for Frontiers in Astronomy and Astrophysics, Beijing Normal University, Beijing 102206, People's Republic of China}
\affiliation{School of Physics and Astronomy, Beijing Normal University, Beijing 100875, People's Republic of China}

\author[0000-0001-9328-4302]{Jun Li (李军)}
\affiliation{Center for Astrophysics, Guangzhou University, Guangzhou 510006, People's Republic of China}

\author[0000-0002-8669-5370]{Dongwei Fan (樊东卫)}
\affiliation{CAS Key Laboratory of Optical Astronomy, National Astronomical Observatories, Chinese Academy of Sciences, Beijing 100101, People's Republic of China}

\author[0000-0003-1218-8699]{Yi Ren (任逸)}
\affiliation{Department of Astronomy, College of Physics and Electronic Engineering, Qilu Normal University, Jinan 250200, People's Republic of China}

\author[0000-0002-9279-2783]{Xiaoxiao Ma (马晓骁)}
\affiliation{CAS Key Laboratory of Optical Astronomy, National Astronomical Observatories, Chinese Academy of Sciences, Beijing 100101, People's Republic of China}
\affiliation{School of Astronomy and Space Science, University of Chinese Academy of Sciences, Beijing 100049, People's Republic of China}

\date{Received 25 June 2024; accepted 24 July 2024}

\begin{abstract}

Intrinsic colors (ICs) of stars are essential for the studies on both stellar physics and dust reddening. In this work, we developed 
an XGBoost model to predict the ICs with the atmospheric parameters $\teff$, $\logg$, and $\meta$. The model was trained and tested 
for three colors at Gaia and 2MASS bands with 1\,040\,446 low-reddening sources. The atmospheric parameters were determined by the 
Gaia DR3 {\gspphot} module and were validated by comparing with APOGEE and LAMOST. We further confirmed that the biases in {\gspphot} 
parameters, especially for $\meta$, do not present a significant impact on the IC prediction. The generalization error of the 
model estimated by the test set is 0.014\,mag for $\BPRPO$, 0.050\,mag for $\BPKO$, and 0.040\,mag for $\JKO$. The model was applied 
to a sample containing 5\,714\,528 reddened stars with stellar parameters from \citet{Andrae2023} to calculate ICs and reddenings. 
The high consistency in the comparison of $\EJK$ between our results and literature values further validates the accuracy of the XGBoost 
model. The variation of $\EBPK/\EBPRP$, a representation of the extinction law, with Galactic longitude is found on large scales.
This work preliminarily presents the feasibility and the accuracy of the machine-learning approach for IC and dust reddening calculation,
whose products could be widely applied to spectro-photometric data. The data sets and trained model can be accessed via 
\url{https://doi.org/10.5281/zenodo.12787594}. The models for more bands will be completed in the following works.
%---------------------------------------------------------------------------------------------------------------------------------
\end{abstract}


\section{Introduction} \label{sec:intro}

Photometric surveys play a vital role in modern astronomy in the discovery and exploration of astronomical objects. For stars, the
intrinsic color (IC), or color index, for different photometric bands, is thus a fundamental and important parameter. On one hand,
it can reveal information about the spectral energy distribution (SED) for the research of stellar physics. On the other hand, IC 
is necessary to estimate the reddening produced by the dust grains and to investigate the extinction laws. For a first attempt, 
\citet{Johnson1966} took the average of the observed colors of nearby stars (within 100\,pc from the Sun) as ICs for various types
of stars for the widely used wide-band filters {\it UBVRIJHKLMN}. Then to overcome the high uncertainty introduced by the limited
sample and the simple averaging, \citet{Ducati2001} invented a new method, based on a much larger catalog, that determined the IC 
for each spectral type by the stars with the bluest colors. This statistical method avoids the confusion of dust reddening in the 
color index. Nevertheless, their bluest stars were selected somehow arbitrarily, and the spectral type is a rough classification 
as well.
%---------------------------------------------------------------------------------------------------------------------------------

Benefiting from modern spectroscopic surveys, such as the Radial Velocity Experiment \citep[RAVE;][]{Steinmetz2006}, the APO 
Galactic Evolution Experiment \citep[APOGEE;][]{Majewski2017}, and the Large Sky Area Multi-Object fiber Spectroscopic Telescope 
\citep[LAMOST;][]{Zhao2012}, the method of \citet{Ducati2001} was further developed for example by \citet{WJ2014} and \citet{Jian2017}. 
With a substantial amount of stars, the zero-reddening ones should constitute the blue edge in the effective temperature versus the 
observed color diagram (e.g. $\teff\ vs.\ \BPRP$). Thus, the median color of the bluest 5\% (some works took 10\%) stars was taken 
to represent the IC for each assigned temperature interval. To further account for the influence of the surface gravity ($\logg$) 
and metallicity ($\meta$) on IC, the star sample was classified as dwarfs and giants according to $\logg$ and separated into different 
$\meta$ groups usually with a step of 0.5\,dex. The relationship between $\teff$ and IC in each sub-sample was then established by 
an analytic function (usually an exponential or polynomial form). This new method was called the ``blue-edge method'' (BEM) and was 
applied in a series of works to build the $\teff$--IC relations from ultraviolet to mid-infrared for the photometric passbands of 
notable surveys \citep[see e.g.][]{Xue2016,Jian2017,Sun2018,WC2019,hz2020}. With the derived IC, the dust reddening can be calculated 
straightforwardly by subtracting the IC from the observed color. Consequently, the BEM was also used in the research on the properties 
of dust grains in the diffuse interstellar medium \citep{Xue2016,Wang2017}, molecular clouds \citep{Cao2023,Li2023}, supernova 
remnants \citep{hz2018,hz2020}, and external Galactic environments \citep[Magellanic clouds;][]{Wang2023}.
%---------------------------------------------------------------------------------------------------------------------------------

The BEM is data-driven and can efficiently deal with large spectro-photometric data sets. Nevertheless, there are some shortcomings 
to this method. First, the ranges of stellar parameters where BEM can derive ICs are limited because BEM needs sufficient number of
stars in 
each $\teff$ interval. For example, hot stars ($\teff\,{>}\,8000$\,K) and metal-poor stars ($\meta\,{<}\,{-}1$\,dex) are almost 
missed in the BEM results. Second, the omitting of the relationship between IC and $\logg$ and $\meta$ in BEM would introduce 
additional uncertainties in modeling IC. Last, the applied analytical function form and the fraction of the selected bluest colors 
are not uniform in different works, which would introduce systematic differences when comparing their results. To overcome these 
shortcomings, we improve the BEM to a machine-learning (ML) approach that models the ICs for various types of stars by the stellar 
atmospheric parameters ($\teff$, $\logg$, $\meta$) using the low-reddening stars as the training and test sets. Similarly, the star-pair
method is also widely used to obtain the ICs and dust reddenings of stars which makes use of stellar parameters and dust-free stars 
\citep[e.g.][]{Yuan2013,ZhangRY2023}. Specifically, for a given star, they selected a set of stars in the control sample 
with similar parameters and low reddenings. The IC was then calculated by a linear fit to the colors of the selected control stars 
and their stellar parameters (see Section 2.5 in \citealt{Yuan2013} for details). Compared to the star-pair method and the BEM, the 
ML approach has two advantages: 1) The ML model directly builds the relationship between ICs and stellar parameters by hyper-parameters, 
with no need for a specific analytical function form and the selection of the control stars. 2) The trained models can promptly predict 
ICs and dust reddenings for new spectro-photometric data, which could be either specifically designed observations in a small sample 
or the next-generation large spectroscopic surveys, such as the Chinese Space Station Telescope \citep[CSST;][]{Zhan2018}, the 
Multi-Object Optical and Near-infrared Spectrograph \citep[MOONS;][]{Cirasuolo2014}, and the 4-meter Multi-Object Spectroscopic 
Telescope \citep[4MOST;][]{deJong2022}. Additionally, the Gaia DR3 data \citep{Vallenari2023} used in this work contains a much 
larger sample of dust-free stars with extensive coverage in the stellar parameter space than those used in previous works. Therefore, 
the IC of stars with extreme parameters (e.g. very hot stars, supergiants, and metal-poor stars) can also be determined. In this work, 
we built an XGBoost model \citep{XGBoost} to estimate stellar ICs from $\teff$, $\logg$, and $\meta$. XGBoost is a scalable end-to-end 
tree-boosting system that is widely used in data science. The training and prediction of the XGBoost model was completed by the 
open-source package {\it xgboost}\footnote{\url{https://github.com/dmlc/xgboost}}.
%--------------------------------------------------------------------------------------------------------------------------------- 

This work focuses on the test and validation of the ML approach by deriving three ICs in optical and near-infrared bands, that is 
$\BPRPO$, $\BPKO$, and $\JKO$, and the corresponding reddenings $\EBPRP$, $\EBPK$, and $\EJK$. The XGBoost models for other bands
will be completed in a follow-up work. The paper is organized as follows: The used data and samples are described and validated in 
Sect. \ref{sect:data}. Section \ref{sect:xgboost} introduces the training of the XGBoost model, the model selection, and the 
performance of the final trained model on the test set. The XGBoost model was applied in Sect. \ref{sect:valid} to an RGB sample 
to derive the dust reddenings. The model was then validated by comparing the reddenings with literature values. We illustrate the 
power of our reddening results in Sect. \ref{sect:cer} by exploring the large-scale variation of the color excess ratio 
$\EBPK/\EBPRP$ which can depict the extinction law and dust properties. Section \ref{sect:summary} summarizes the main results of 
this work.

\section{Data} \label{sect:data}

\subsection{Gaia DR3 and 2MASS} \label{subsect:gaia-2mass}

Besides the widely used astrometry and broad-band photometry, Gaia DR3 includes extensive information on the astrophysical properties 
of stars \citep[see e.g.][]{Creevey2023_GS,Recio-Blanco2023,Kordopatis2023,GSP-phot}. Among them, \citet{GSP-phot} produced a homogeneous 
catalog of stellar parameters, including $\teff$, $\logg$, $\meta$, absolute $G$-band magnitude, radius, distance, and extinction, 
for 471 million sources with $G\,{<}\,19$\,mag (published as part of Gaia DR3), derived by the General Stellar Parameterizer from 
Photometry ({\gspphot}) module of the Astrophysical parameters inference system \citep[Apsis;][]{Creevey2023}, based on the analysis 
of their astrometry, photometry, and low-resolution BP/RP spectra \citep{DeAngeli2023}. The Gaia DR3 astrometry, photometry, and 
{\gspphot} parameters can be accessed through the datalink interface of the Gaia Archive\footnote{\url{https://gea.esac.esa.int/archive/}}. 
%---------------------------------------------------------------------------------------------------------------------------------

We made use of the parallax, $G_{\rm BP}$ and $G_{\rm RP}$ photometry, and the {\gspphot} stellar atmospheric parameters ($\teff$, 
$\logg$, $\meta$) in this work, as well as the {\it J} ($1.25\,\micron$) and $\Ks$ ($2.16\,\micron$) bands photometric data from 
the 2MASS point-source catalog \citep{Skrutskie2006}. The 2MASS catalog was cross-matched with the Gaia DR3 main catalog within 
$1\arcsec$, resulting in 208 million common objects. 
%---------------------------------------------------------------------------------------------------------------------------------

Due to the extremely low resolution of BP/RP spectra (20--60 for BP, 30--50 for RP, \citealt{Montegriffo2023}), there are two important 
shortcomings of the {\gspphot} parameters (see Section 3.5 in \citealt{GSP-phot} for detailed discussions): 1) BP/RP spectra are 
simply not informative on $\meta$ which results in strong systematic differences and significant outliers in external validation. 
2) {\gspphot} overestimates both $\teff$ and extinction for sources with $A_0\,{\gtrsim}\,2$\,mag\footnote{$A_0$ is the monochromatic 
extinction at 541.4\,nm} due to the temperature--extinction degeneracy. The latter shortcoming does not affect our task as the XGBoost 
model is trained by low-reddening stars. For the former one, we discuss the possible influence of the {\gspphot} parameter biases 
on the determination of the ICs in Sect. \ref{subsect:gspphot-bias}.
%---------------------------------------------------------------------------------------------------------------------------------

\subsection{Building training and test sets} \label{subsect:samples}

To model the ICs with stellar atmospheric parameters, we need to first select a sample of stars with low reddenings and high-quality
parameters and photometry. The sample was constructed by the following requirements:
%---------------------------------------------------------------------------------------------------------------------------------

\begin{enumerate}
    \item Sources need to have {\gspphot} determinations of $\teff$, $\logg$, and $\meta$. 
    \item Sources should have small reddenings: $\EBV$ from \citet{Planck2016dust} is smaller than 0.02\,mag.
    \item The error of parallax is smaller than 30\% as {\gspphot} performs better for objects with high-quality parallax.
    \item The photometric error is smaller than 0.02\,mag in $G_{\rm BP}$ and $G_{\rm RP}$ bands and smaller than 0.05\,mag in {\it J} 
    and $\Ks$ bands.
    \item Sources with very large uncertainties of {\gspphot} parameters, exceeding 95\% of the total, are removed.
\end{enumerate}
%---------------------------------------------------------------------------------------------------------------------------------

\begin{figure*}[ht!]
  \centering
  \includegraphics[width=8cm]{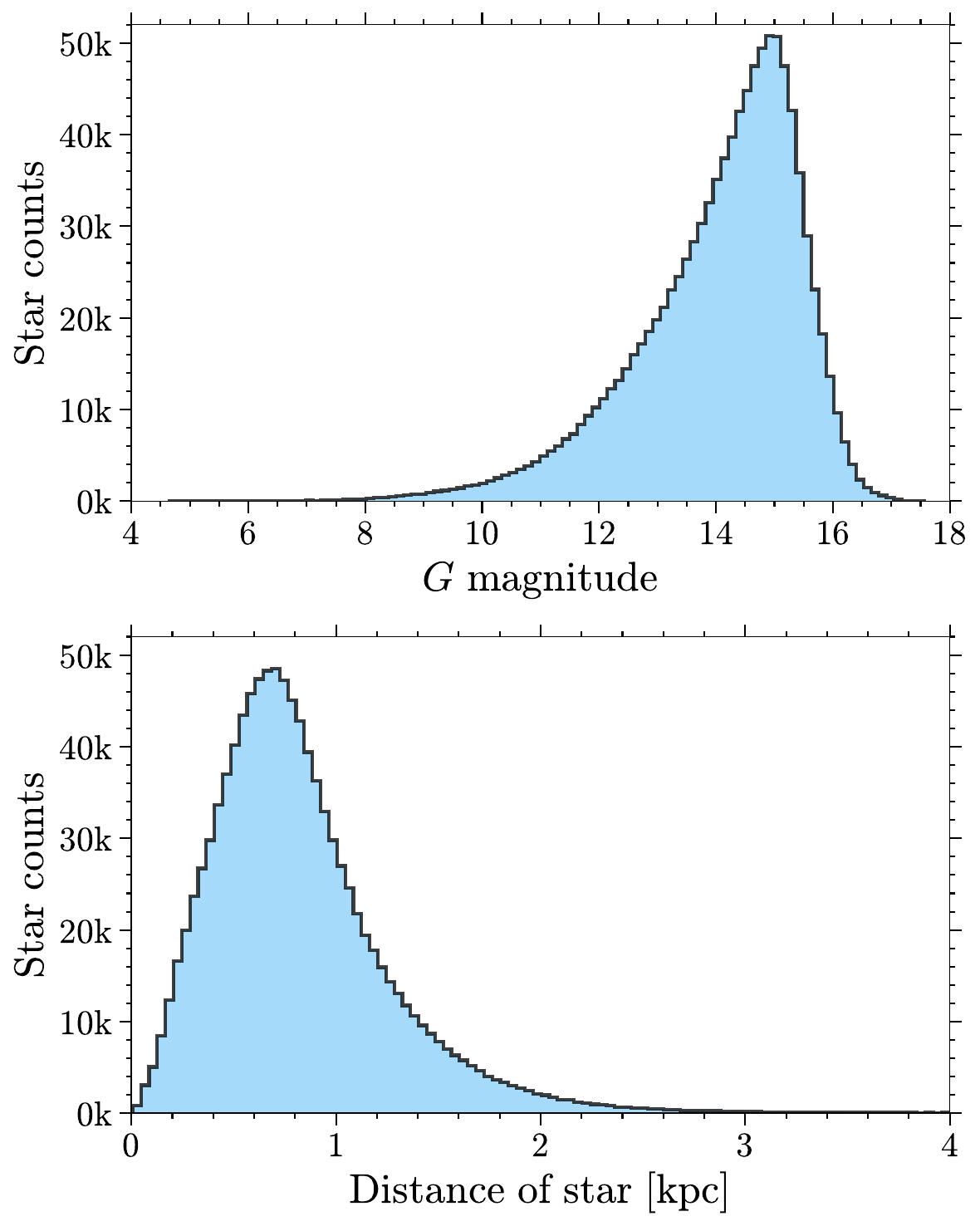}
  \caption{Distribution of the Gaia $G$--band magnitude ({\it top panel}) and the distance (inverse of parallax, {\it bottom panel}) 
  of the 1\,040\,446 selected stars with low reddenings.}
  \label{fig:G-dist}
\end{figure*}

The final selected sample contains 1\,040\,446 sources, of which the distribution of $G$--band magnitude and distance (inverse of
the parallax) is shown in Figure \ref{fig:G-dist}. The peak of $G$ magnitude is around 15\,mag, and the dramatic decrease in number 
after it is due to the low-quality photometry for fainter stars. The sample is mainly consistent with nearby stars of which 95\% 
are within 1.7\,kpc. This sample was further randomly separated into the training set containing 832\,356 sources (80\%) and the 
test set containing 208\,090 sources (20\%).
%---------------------------------------------------------------------------------------------------------------------------------

\begin{figure*}[ht!]
  \centering
  \includegraphics[width=10cm]{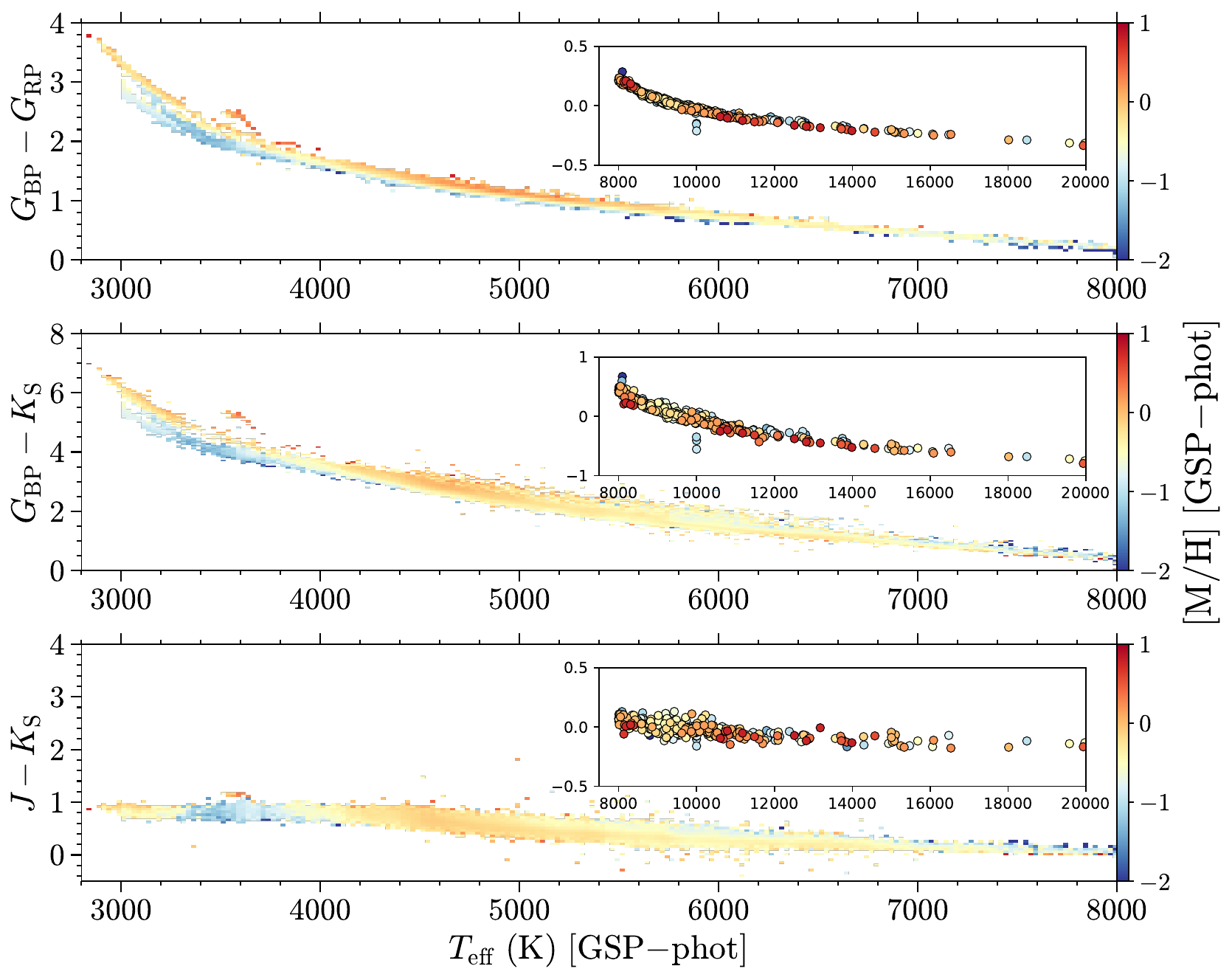}
  \caption{Correlation between $\teff$ and observed colors for the 1\,040\,446 selected stars with low reddenings. The color-coded 
  map presents the average $\meta$ in each 25\,K\,$\times$\,0.05\,mag bin. The zoom-in panels show the correlation for hotter stars 
  up to 20\,000\,K, color-coded by their $\meta$. }
  \label{fig:teff-color}
\end{figure*}

Figure \ref{fig:teff-color} shows the monochromatic correlation between $\teff$ and observed colors for the whole selected sample
from $\teff\,{\sim}\,3000$\,K up to as high as 20\,000\,K. The span of the $\teff$--color curve would present the intrinsic broadening
of their relationship, plus the additional effect of $\logg$ and $\meta$, as well as the parameter/photometry uncertainty. The colored 
map in Fig. \ref{fig:teff-color} shows an apparent influence of $\meta$ on the ICs which was discussed before for example in 
\citet{Jian2017} and \citet{hz2020}. This influence is strong in optical bands and becomes weaker in infrared. $\JKO$ is not
sensitive to $\meta$ and its colored map illustrates the main part of $\meta$ in the corresponding $\teff$ range of the selected sample.
For instance, 3200--3800\,K is dominated by stars with $\meta$ between --1 and --0.5\,dex and $\meta$ in the interval 4000--5000\,K 
is mainly around zero.
%---------------------------------------------------------------------------------------------------------------------------------

\subsection{Validate {\gspphot} stellar parameters} \label{subsect:data-check}

\begin{figure*}[ht!]
  \centering
  \includegraphics[width=16cm]{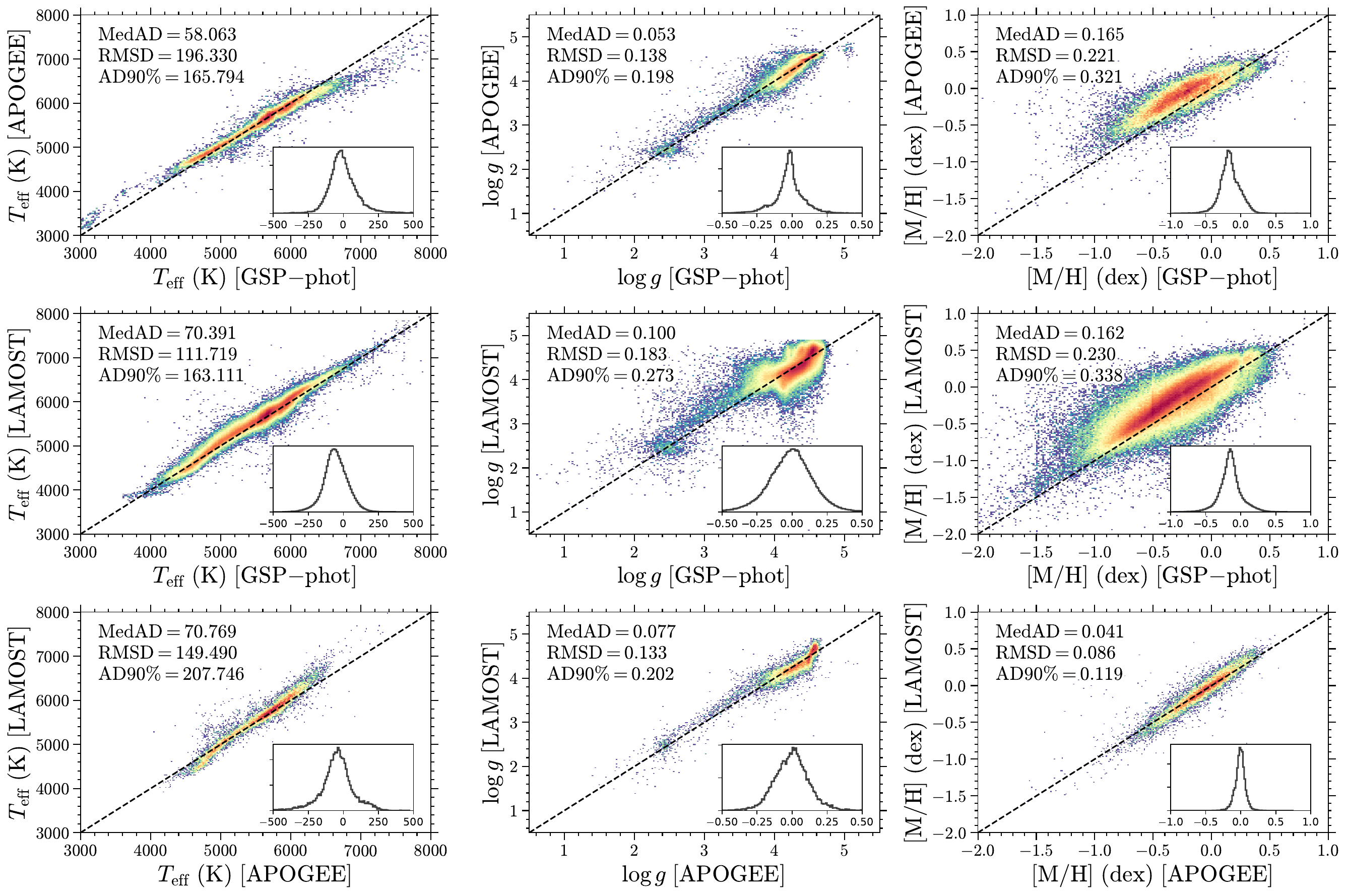}
  \caption{Comparison of the stellar atmospheric parameters between the common sources from {\gspphot}, APOGEE, and LAMOST. The
  color in each panel represents the number density of stars calculated in each bin with a step of 25\,K for $\teff$, 0.025 for
  $\logg$, and 0.015\,dex for $\meta$. The dashed black lines show the one-to-one correspondence. The zoom-in panels present the
  distribution of the parameter differences ({\gspphot}\,--\,APOGEE/LAMOST). The marked statistics are the median absolute difference 
  (MedAD), the root-mean-square difference (RMSD), and the absolute difference not exceeded by 90\% of sources (AD90\%).}
  \label{fig:check-sample}
\end{figure*}

To validate the accuracy of the {\gspphot} stellar parameters, we cross-matched the selected sample with the catalogs of 
APOGEE--DR17\footnote{\url{https://www.sdss4.org/dr17/irspec/spectro_data/\#SpectralData}} and LAMOST--DR10\footnote{\url{http://www.lamost.org/dr10/}}
and got 30\,130 and 243\,148 common sources, respectively. Additionally, there are 14\,993 sources in the selected sample which were
observed by both APOGEE and LAMOST. Figure \ref{fig:check-sample} shows the comparison of the stellar parameters\footnote{We note 
that for the metallicity, we used $\meta$ for {\gspphot} and APOGEE, and $\rm [Fe/H]$ for LAMOST.} for the common sources, together 
with the marks of the statistics of the absolute differences: the median absolute difference (MedAD), the root-mean-square 
difference (RMSD), and the absolute difference not exceeded by 90\% of sources (AD90\%).
%---------------------------------------------------------------------------------------------------------------------------------

For $\teff$, {\gspphot} is in good mutual agreement with both APOGEE and LAMOST, where over 90\% of the common sources have $\teff$
differences smaller than 170\,K. Their statistics are similar to those between APOGEE and LAMOST and dramatically reduce compared
to the comparison in \citet{GSP-phot} with a total sample containing highly reddened sources (see Sect. 3.5.1 and Table 1 in 
\citealt{GSP-phot} for details). There is a deviation between {\gspphot} and APOGEE for stars hotter than $\sim$6500\,K, which would
be a problem in APOGEE estimates as this deviation is seen in APOGEE--LAMOST comparison as well but not visible in {\gspphot}--LAMOST
comparison. The high quality of the {\gspphot} $\logg$ estimates for sources with good parallaxes has been illustrated in \citet{GSP-phot}.
The MedAD of the {\gspphot}--APOGEE differences for $\logg$ significantly reduces from 0.218 in \citet{GSP-phot} to 0.053 in our sample, 
as well for RMSD and AD90\%. The statistics also slightly decrease for the {\gspphot}--LAMOST comparison. We note that the distribution 
of the $\logg$ differences between {\gspphot} and APOGEE is even more concentrated to zero than that between APOGEE and LAMOST. 
Although the statistics of the absolute differences for $\meta$ comparison decreased by nearly a factor of two from \citet{GSP-phot} 
to our sample, there is still a systematic underestimation of {\gspphot} $\meta$ about 0.16\,dex. Nevertheless, in Sect. 
\ref{subsect:gspphot-bias}, we will show that these {\gspphot} parameter biases do not have an apparent impact on the determination 
of the ICs.
%---------------------------------------------------------------------------------------------------------------------------------

\section{Training XGBoost} \label{sect:xgboost} 

\subsection{Training and cross-validation} \label{subsect:model}

The input features for XGBoost are {\gspphot} stellar parameters, that is $\teff$, $\logg$, and $\meta$. And the predicted labels
are three ICs, that is $\BPRPO$, $\BPKO$, and $\JKO$. For low-reddening sources in the training and test sets, their observed colors 
were taken as the intrinsic ones. We trained the XGBoost model by optimizing six hyper-parameters, leaving others as default in 
{\it xgboost}:

\begin{enumerate}
    \item `max\_depth': Maximum tree depth for base learners.
    \item `n\_estimators': Number of gradient-boosted trees. 
    \item `learning\_rate': Boosting learning rate.
    \item `gamma': Minimum loss reduction required to make a further partition on a leaf node of the tree.
    \item `reg\_lambda': L2 regularization term on weights.
    \item `max\_leaves': Maximum number of leaves.
\end{enumerate}
%---------------------------------------------------------------------------------------------------------------------------------

The initial guesses of the hyper-parameters were arbitrarily selected. Then we trained the model with a set of values for one 
hyper-parameter, keeping others unchanged. A 10-fold cross-validation\footnote{The training set was randomly separated into ten 
equal groups. In each cross-validation run, one group was treated as the validation set with the rest nine groups as the training
set.} was applied to choose the best value for this hyper-parameter. If one hyper-parameter gets a new value, the optimization
process will be repeated for all the others until no hyper-parameters change. The best combination of the hyper-parameters is
shown below. See Appendix \ref{app:cv} for more details and Fig. \ref{fig:cv-curve} for the validation curves of these 
hyper-parameters.
%---------------------------------------------------------------------------------------------------------------------------------

\begin{verbatim}
max_depth = 6
n_estimators = 300
learning_rate = 0.1
gamma = 0.02
reg_lambda = 4
max_leaves = 10
\end{verbatim}
%---------------------------------------------------------------------------------------------------------------------------------

The XGBoost model was then finally trained with the best hyper-parameters and the full training set. 
We attach a Python script in Appendix \ref{app:script} showing how to use the XGBoost model to predict ICs for given parameters.
Moreover, to facilitate the usage of our results, we build an IC catalog containing a representative set of ICs as a function of
typical $\teff$, $\logg$, and $\meta$. Due to the highly inhomogeneous star distribution in atmospheric parameter space and the
limited coverage of our reference sample, the selection of parameter combinations is not uniform but with a simple filtering method.
Specifically, we first generated a set of parameter grids: 3000\,K to 8000\,K with a step of 200\,K for $\teff$ plus 9000\,K to
20\,000\,K with a step of 1000\,K, 1 to 5 with a step of 0.2 for $\logg$, and --2 to 0.6\,dex with a step of 0.2\,dex for $\meta$.
Then we counted the number of reference stars within the step size of each grid point and only kept the parameter combinations if
we could find reference stars therein. For example, for $\teff\,{=}\,5000$\,K, $\logg\,{=}\,4$, and $\meta\,{=}\,0.2$\,dex, there
are 23 reference stars within 4900--5100\,K, 3.9--4.1, and 0.1--0.3\,dex. The IC catalog finally contains 1195 combinations of
$\teff$, $\logg$, and $\meta$, which could roughly describe the applicable scope of the XGBoost model.
The trained model, the data sets, the IC catalog, and the script are published online at \url{https://doi.org/10.5281/zenodo.12787594}. 
A subset of the IC catalog is listed in Table \ref{tab:IC} for a quick look, which contains all the valid $\teff$ grids
under the combinations of $\logg\,{=}\,1$, 2, 3, 4, 5 and $\meta\,{=}\,{-}2$, --1, 0, 0.4\,dex. Over 65\% of the $\teff$ grids are 
for dwarfs with $\logg\,{=}\,4$. The ICs at typical $\teff$ are consistent with the determinations in \citet[][Table 1]{WJ2014} 
and \citet[][Tables 7 and 9]{Jian2017}. 
%---------------------------------------------------------------------------------------------------------------------------------

\subsection{Performance on test set} \label{subsect:test_set}

\begin{figure*}[ht!]
  \centering
  \includegraphics[width=16cm]{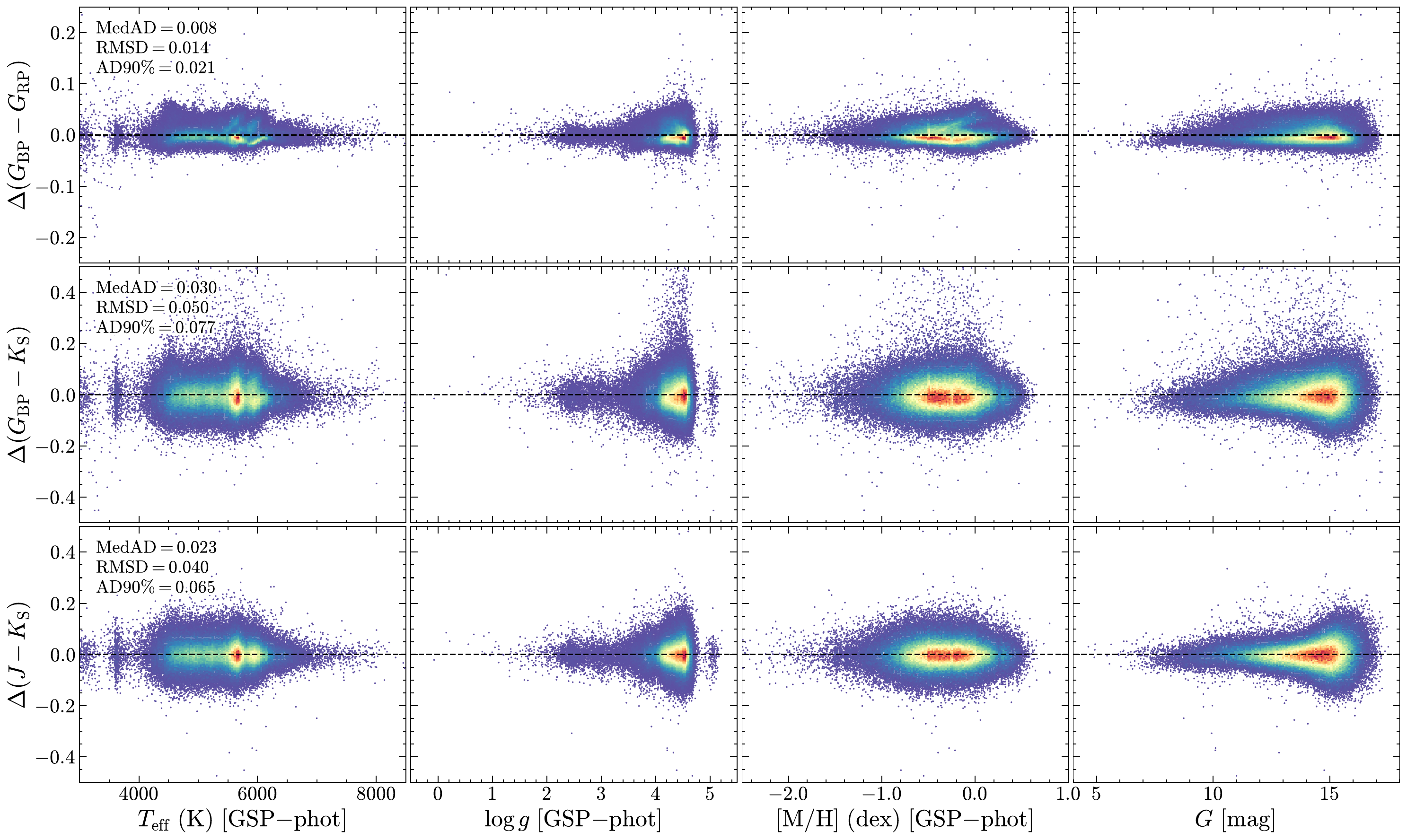}
  \caption{Differences between observed and XGBoost predicted colors for the test set, as a function of the {\gspphot} stellar 
  parameters ($\teff$, $\logg$, and $\meta$) and $G$--magnitude. The color-coded map represents the number density of the sources 
  on a linear scale. The median absolute difference (MedAD), the root-mean-square difference (RMSD), and the absolute difference 
  not exceeded by 90\% of sources (AD90\%) for each color are also marked in the first column of each panel.}
  \label{fig:test-set}
\end{figure*}

The trained XGBoost model was applied to the test set (208\,090) to evaluate its performance. Figure \ref{fig:test-set} shows the
differences between the observed and predicted colors, as a function of stellar parameters and $G$--magnitude. For most of the 
sources, the XGBoost predictions are accurate, with AD90\% as 0.021\,mag for $\Delta (\BPRP)$, 0.077\,mag for $\Delta (\BPK)$, and 
0.065\,mag for $\Delta (\JK)$. There is no significant bias even for the parameter regions with only a few stars, such as 
$\teff\,{>}\,7000$\,K, $\logg\,{<}\,2$, and $\meta\,{<}\,{-}1.5$\,dex. There are only a few outliers, which are most prominent for 
$\BPK$. These outliers concentrated around $\logg\,{\sim}\,4.4$ but were not sensitive to $\teff$ and $\meta$, which would indicate 
a problematic estimate of the $\logg$ of these sources. The structural differences, mainly related to $\teff$ and $\meta$, are more
prominent for $\BPRP$ than the other two colors because of the much better photometric quality of Gaia than 2MASS. These differences
may be caused by the bias of the {\gspphot} parameters (see discussions in Sect. \ref{subsect:gspphot-bias}) but are mainly within 
0.03\,mag.
%---------------------------------------------------------------------------------------------------------------------------------

Illustrated by Fig. \ref{fig:test-set}, it seems that the color differences are not sensitive to $\teff$ and $\meta$ but increase 
with $\logg$ and $G$--magnitude. Meanwhile, the number density of sources is also uniform for $\teff$ and $\meta$ but increases for 
$\logg$ and $G$, which would create a visual impact on the distribution of the color differences. Therefore, we recalculated MedAD, 
RMSD, and AD90\% in different bins to explore the sensitivity of the color differences to stellar parameters and $G$--magnitude. The 
results are presented in Fig. \ref{fig:test-set-sub}. Despite the extreme values and dramatic fluctuations caused by the insufficient 
numbers of sources in some bins, MedAD, RMSD, and AD90\% of all three colors vary with stellar parameters and $G$ in a similar manner: 
these statistics generally decrease with $\teff$ and $\meta$ but increase with $\logg$ and $G$. Nevertheless, their variations are 
generally smaller than their values. Further, the systematic variations of $\BPK$ and $\JK$ are more apparent than $\BPRP$. 
%---------------------------------------------------------------------------------------------------------------------------------

Considering RMSD as an estimate of the generalization error, the uncertainty of the XGBoost prediction for ICs, 0.014\,mag for $\BPRPO$, 
0.050\,mag for $\BPKO$, and 0.040\,mag for $\JKO$, is on the same order of magnitude as the constraints for the photometric errors 
of Gaia and 2MASS bands (0.02\,mag and 0.05\,mag), which implies that the accuracy of the XGBoost prediction has reached the limit 
of the observational data. RMSD would be a conservative estimate and the XGBoost uncertainty should slightly vary for different 
types of stars, as well as their brightness, as shown in Fig. \ref{fig:test-set-sub}.
%---------------------------------------------------------------------------------------------------------------------------------

The color differences of hotter stars between 8000\,K and 20\,000\,K, shown in Fig. \ref{fig:test-set-hot}, are also not sensitive to 
stellar parameters and $G$--magnitudes. Despite a few outliers, 90\% of the differences are within 0.026\,mag for $\BPRP$, 0.088\,mag
for $\BPK$, and 0.055\,mag for $\JK$, which verifies that the XGBoost model has a quasi-uniform performance for extensive types of
stars.
%---------------------------------------------------------------------------------------------------------------------------------

\subsection{Effect of the {\gspphot} biases} \label{subsect:gspphot-bias}

\begin{figure*}[ht!]
  \centering
  \includegraphics[width=16cm]{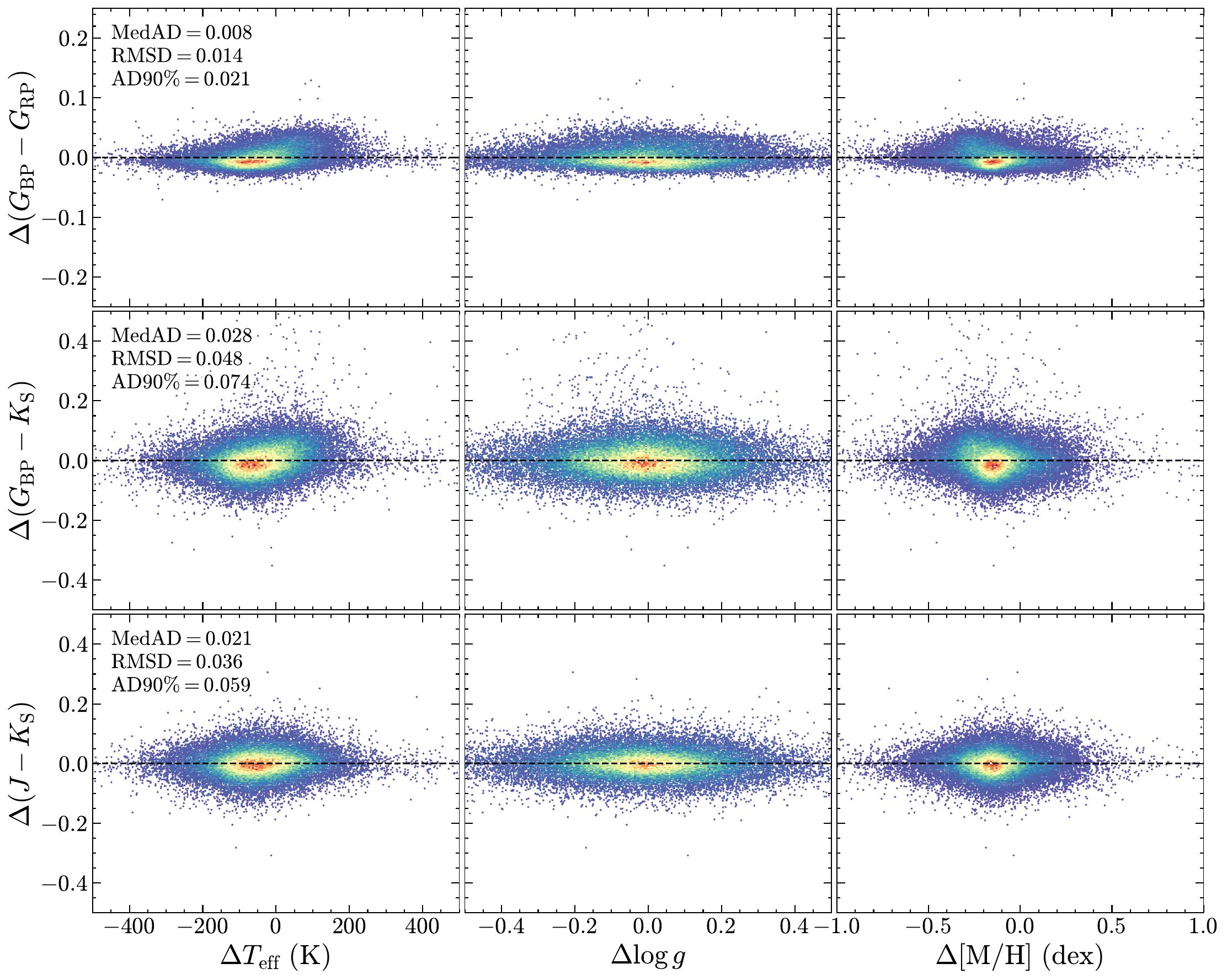}
  \caption{Differences between observed and XGBoost predicted colors as a function of the parameter differences ({\gspphot} --
  APOGEE/LAMOST) for 51\,417 sources selected from the test set. The color-coded map represents the number density of the sources 
  on a linear scale. The median absolute difference (MedAD), the root-mean-square difference (RMSD), and the absolute difference 
  not exceeded by 90\% of sources (AD90\%) for each color are also marked in the first column of each panel.}
  \label{fig:test-EvP}
\end{figure*}

To investigate the impact of the {\gspphot} parameter biases, especially $\meta$, on the XGBoost model, we collected from the test
set 6033 sources with APOGEE parameters and 48\,382 sources with LAMOST parameters. After excluding duplicated sources (keep using
APOGEE parameters), we compare the color differences (observed color -- XGBoost predictions) and the parameter differences ({\gspphot}
-- APOGEE/LAMOST) for 51\,417 sources shown in Fig. \ref{fig:test-EvP}. The MedAD, RMSD, and AD90\% of the color differences for 
this subsample are very close to those of the whole test set. $\Delta (\JK)$ is not sensitive to all the parameter differences. 
$\Delta (\BPRP)$ and $\Delta (\BPK)$ are also insensitive to $\Delta \logg$ and the main part of $\Delta \teff$ and $\Delta \meta$. 
But there exists a subbranch in $\Delta \teff$ and $\Delta \meta$ where $\Delta (\BPRP)$ and $\Delta (\BPK)$ increase with the 
absolute parameter differences. This could be a piece of evidence for the impact of the {\gspphot} parameter biases on the IC determination. 
Nevertheless, we note that 1) this impact has a similar degree to that of the intrinsic uncertainty of the model and the photometric 
errors. 2) The systematic deviation occurs for a special type of star and does not always keep growing with the parameter differences. 
For instance, the color difference concentrated to zero with a very small dispersion for $|\Delta \teff|\,{\gtrsim}\,200$\,K. Another 
interesting hint is that the negative mean of the color differences seen in Figs. \ref{fig:test-set} and \ref{fig:test-EvP} would 
be related to the negative mean of $\Delta \teff$, $\Delta \logg$, and $\Delta \meta$.
%---------------------------------------------------------------------------------------------------------------------------------

\section{External validation: dust reddening} \label{sect:valid}

In this section, we validate the XGBoost model by applying it to a sample of reddened stars to calculate the dust reddening and 
comparing the results with literature values.
%---------------------------------------------------------------------------------------------------------------------------------

\subsection{Sample selection} \label{subsect:ev-sample}

Due to the temperature--extinction degeneracy, the {\gspphot} parameters were no longer used for reddened stars. Several efforts have 
been made to re-estimate or calibrate stellar parameters from BP/RP spectra trained on external data sets \citep[e.g.][]{Rix2022,
Andrae2023,ZhangXY2023,Avdeeva2024}. The products of two of them were used in this work.
%---------------------------------------------------------------------------------------------------------------------------------

One is \citet[][hereafter A23]{Andrae2023}\footnote{Data access: \url{https://doi.org/10.5281/zenodo.7945154}} who provided improved 
$\teff$, $\logg$, and $\meta$ for $\sim$175 million Gaia DR3 sources, trained on APOGEE data set. They got high accuracy in the 
prediction of stellar parameters but not for OBA-type stars due to the limited temperature range of the APOGEE data. Thus, 
\citetalias{Andrae2023} selected $\sim$17.5 million high-purity RGB stars from their results with high-quality parameter estimates
and low contamination of hot stars (see Sect. 4.2 in \citetalias{Andrae2023} for details). Our XGBoost model predicted the intrinsic
colors, $\BPRPO$, $\BPKO$, and $\JKO$, for sources in the A23-RGB sample with the stellar parameters from \citetalias{Andrae2023}.
The reddening in corresponding bands, that is $\EBPRP$, $\EBPK$, and $\EJK$, were then calculated straightforwardly by subtracting 
the observed colors.
%---------------------------------------------------------------------------------------------------------------------------------

The other work is \citet[][hereafter Z23]{ZhangXY2023}\footnote{Data access: \url{https://doi.org/10.5281/zenodo.7811871}} who built an 
empirical forward model from the LAMOST training sample to estimate the stellar atmospheric parameters, distances, and extinctions 
for 220 million stars with BP/RP spectra in Gaia DR3. The estimates of stellar parameters of \citetalias{ZhangXY2023} and 
\citetalias{Andrae2023} generally agree with each other (see Sect. 5.3 and Fig. 17 in \citetalias{ZhangXY2023}) with systematic 
differences in $\logg$ and $\meta$ primarily due to their different training sets. We made use of $\EJK$\footnote{$\EJK$ was 
calculated by the extinction scalar ($E$) and vector coefficients at $J$ and $\Ks$ bands given in \citetalias{ZhangXY2023}.} from 
\citetalias{ZhangXY2023} as literature values for the reddening comparison.  
%---------------------------------------------------------------------------------------------------------------------------------

We cross-matched the A23-RGB sample and \citetalias{ZhangXY2023} catalog and got $\sim$11 million common stars. We further
derived $\EJK$\footnote{See Table 1 in \citet{Green2019} for the reddening vector coefficients.} from the dust map of \citet[][
hereafter G19]{Green2019} using the positions and distances (inverse of parallax) of the common stars via the Python package {\it 
dustmaps} \citep{Green2018python}. Only 6.7 million stars got valid $\EJK$ from the dust map due to its limited sky coverage.
{\gspphot} also gave $\EBPRP$ measurements for these RGB stars. We converted $\EBPRP$ to $\EJK$ with a simple coefficient (0.3651)
from the extinction law of \citet{CCM89} to have a uniform comparison with literature values.
%---------------------------------------------------------------------------------------------------------------------------------

\subsection{Selecting training-like sources} \label{subsect:ev-qc}

\begin{figure*}[ht!]
  \centering
  \includegraphics[width=16cm]{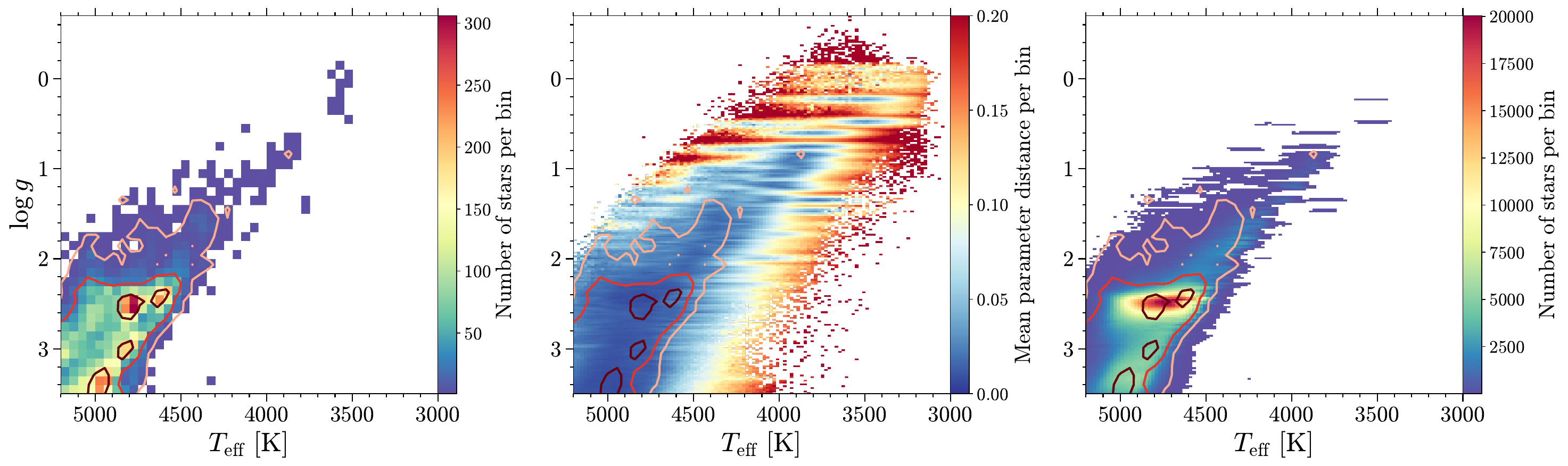}
  \caption{Kiel diagram for the training set (left) and the A23-RGB sample selected from \citet[][middle and right]{Andrae2023}. 
  The left panel is colored by the number of stars in each 50\,K\,$\times$\,0.1 bin. The contours in each panel from light red to
  dark red represent 1\%, 10\%, and 50\% of the maximum of the star number (306), respectively. The middle panel shows the mean 
  parameter distance of stars in the A23-RGB sample to the training set (see Sect. \ref{subsect:ev-qc}) in each 20\,K\,$\times$\,0.02 
  bin. The right panel contains 9 million stars in the A23-RGB sample filtered by parameter distance smaller than 0.03 and is 
  colored by the star number in each 20\,K\,$\times$\,0.02 bin.}
  \label{fig:para-dist}
\end{figure*}

Although ML algorithms can proficiently learn input features in the training set, a widely known shortcoming is that the performance
of the ML model would rapidly degrade when the target sample does not share common features with the training set. For example, the
A23-RGB sample contains a set of cool giants/supergiants ($\teff\,{\lesssim}\,4000$\,K and $\logg\,{\lesssim}\,1$) but these types 
of stars are very few in our training set, which would lead to systematics in the determination of their ICs. One solution is to 
classify the target sample into `training-like' and `training-unlike' sources. For the studies on spectral data, the dimensionality 
reduction technique (e.g. t-SNE) was usually applied to the spectra to provide a two-dimensional (2D) map where the geometric distance 
of each target source to the training set can be calculated. Then the training-like sources in the target sample can be selected 
by limiting such distance. Details can be found in \citet{Ambrosch2023} and \citet{Guiglion2024} for example. In our task, as the 
stellar SED can be largely characterized by $\teff$, $\logg$, and $\meta$, we calculated the parameter distance of each target 
source to its closest neighbor in the training set defined as
%---------------------------------------------------------------------------------------------------------------------------------

\begin{equation}
    d_P = \left[\sum_{i=1}^{3} \left( \frac{P_{\rm target}^{i} - P_{\rm train}^{i}}{P_{\rm target}^{i}} \right)^2\right]^{1/2},
\end{equation}
%---------------------------------------------------------------------------------------------------------------------------------

\noindent where $\mathbi{P}\,{=}\,\{\teff,\,\logg,\,\meta\}$. Figure \ref{fig:para-dist} illustrates the distribution of $d_P$ in
the Kiel diagram (middle panel). Referring to the number density distribution of the training set (left panel), $d_P$ is mainly
under 0.03 in the high-density regions and dramatically increases in poorly covered regions. We applied an upper limit of 0.03 for 
$d_P$, which corresponds to 2\% difference on average of the stellar parameters. This limit resulted in 9 million training-like
sources in the A23-RGB sample, whose number density distribution is shown in the right panel in Fig. \ref{fig:para-dist}. The main
part of the filtered A23-RGB sample is highly consistent with the high-density region of the training set. Besides the $d_P$ cut,
we applied the basic cut and the confidence cut of stellar parameters given by \citetalias{ZhangXY2023} (see Table 2 and Sect. 4.4
therein) as well to have a reliable \citetalias{ZhangXY2023} $\EJK$ measurements. The final target sample contains 5\,714\,528 
sources. A shortcoming of $d_P$ cut would be that we assumed an equal weight for $\teff$, $\logg$, and $\meta$, but these parameters
would have different impacts on the stellar SED, resulting in a possibility that two stars with small $d_P$ would have very different 
physical properties. Future improved methods could consider further constraints on the density of the training set or adjustments 
of the parameter weights.
%---------------------------------------------------------------------------------------------------------------------------------

\subsection{Reddening comparison} \label{subsect:ev-ext}

\begin{figure*}[ht!]
  \centering
  \includegraphics[width=14cm]{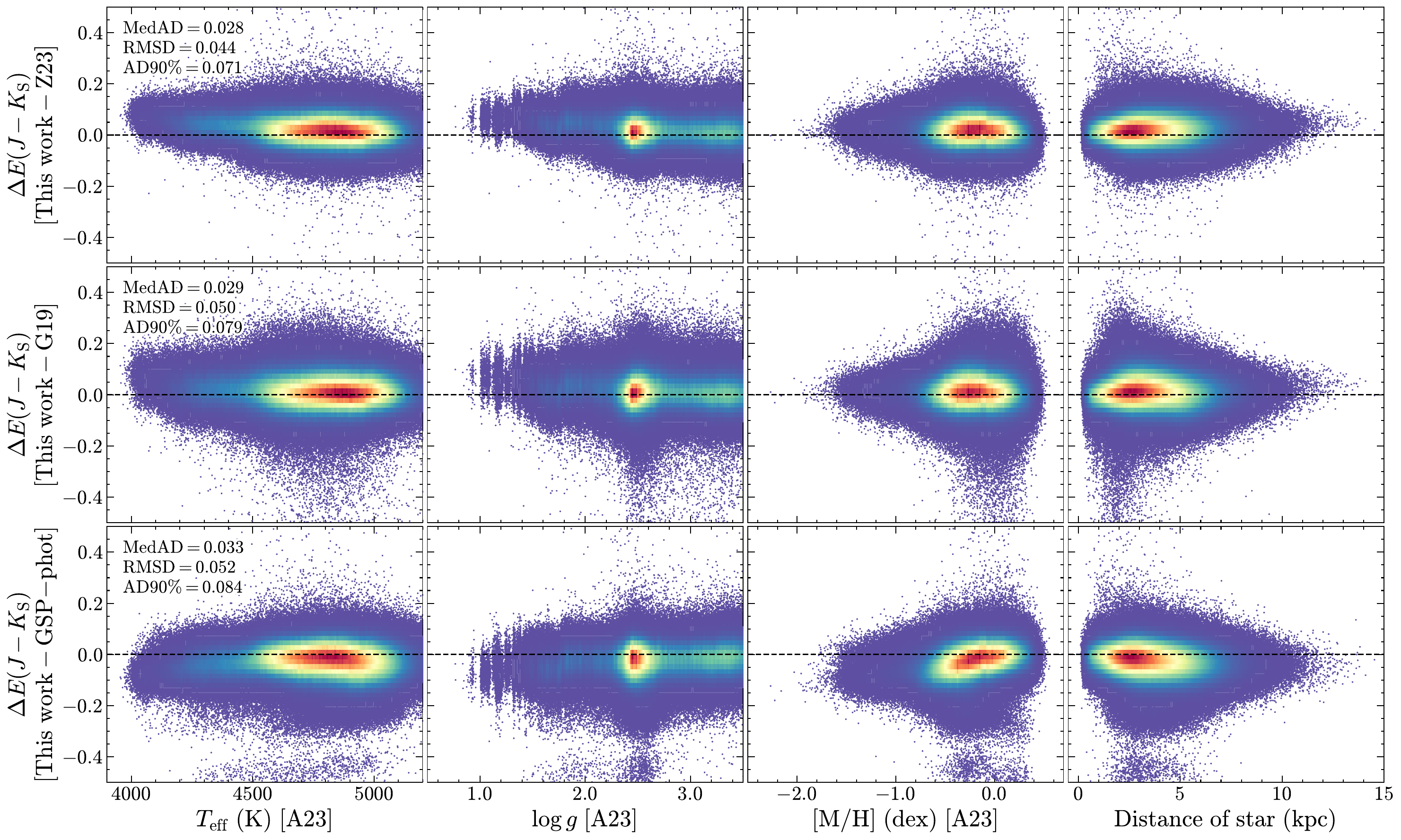}
  \caption{Differences of $\EJK$ between this work and literature values from \citet[][top]{ZhangXY2023}, \citet[][middle]{Green2019} 
  and \citet[][bottom]{GSP-phot} as a function of stellar atmospheric parameters from \citet{Andrae2023} and distance (inverse of 
  parallax) for 5\,714\,528 sources in the target sample. The color-coded map represents the number density of the sources on a 
  linear scale. The median absolute difference (MedAD), the root-mean-square difference (RMSD), and the absolute difference not 
  exceeded by 90\% of sources (AD90\%) for the $\EJK$ difference are also marked in the first column of each panel.}
  \label{fig:reddening}
\end{figure*}

The differences of $\EJK$ between this work and literature values are shown in Fig. \ref{fig:reddening} as a function of the stellar
parameters from \citetalias{Andrae2023} and the distance (inverse of parallax). $\EJK$ derived by our XGBoost model is highly
consistent with literature values, with MedAD, RMSD, and AD90\% of $\Delta \EJK$ very close to (only slightly larger than) those 
of the test set, with RMSD as 0.044\,mag for the comparison with \citetalias{ZhangXY2023}, 0.050\,mag with \citetalias{Green2019},
and 0.052\,mag with {\gspphot}, which implies that the color differences are dominated by the 2MASS photometric errors rather than 
the intrinsic uncertainty between different methods. $\EJK$ calculated from our XGBoost model is overall systematically larger
than that from \citetalias{Green2019} and \citetalias{ZhangXY2023}, but smaller than that of {\gspphot}. A significant overestimation 
of $\EJK$ (${\lesssim}\,{-}0.4$) can be clearly seen for {\gspphot} and \citetalias{Green2019} but only accounts for a very small 
fraction because most of the sources in the target sample are not highly reddened (only 5.6\% have $A_0\,{>}\,2$\,mag). Such 
overestimation was efficiently avoided in our results and \citetalias{ZhangXY2023} for much fewer outliers seen in their comparison. 
%---------------------------------------------------------------------------------------------------------------------------------

The $\EJK$ difference performs in a similar manner with \citetalias{ZhangXY2023} and \citetalias{Green2019}. For the main part, 
$\Delta \EJK$ is not sensitive to stellar parameters and distance and presents a larger positive offset for \citetalias{ZhangXY2023} 
than that for \citetalias{Green2019}. The systematic overestimation of our results found at low $\teff$, low $\logg$, and large 
distance correspond to distant cool giants which are not well covered in our training set. As IC generally decreases with 
$\teff$ (see Fig. \ref{fig:teff-color}), the extrapolated prediction of the XGBoost model based on hotter stars would underestimate 
the IC of cool giants, resulting in an overestimation of $\EJK$. For $\meta$, $\Delta \EJK$ keeps around zero down to --2\,dex, which 
further confirms that our XGBoost model trained on low-reddening sources is not affected by the {\gspphot} bias. On the contrary, 
{\gspphot} predicted an even larger $\EJK$ compared to our results for these distant cool giants. Moreover, $\EJK$ was systematically 
overestimated by {\gspphot} for metal-poor stars as well.  
%---------------------------------------------------------------------------------------------------------------------------------

\section{Application: distribution of color excess ratio} \label{sect:cer}

\begin{figure*}[ht!]
  \centering
  \includegraphics[width=12cm]{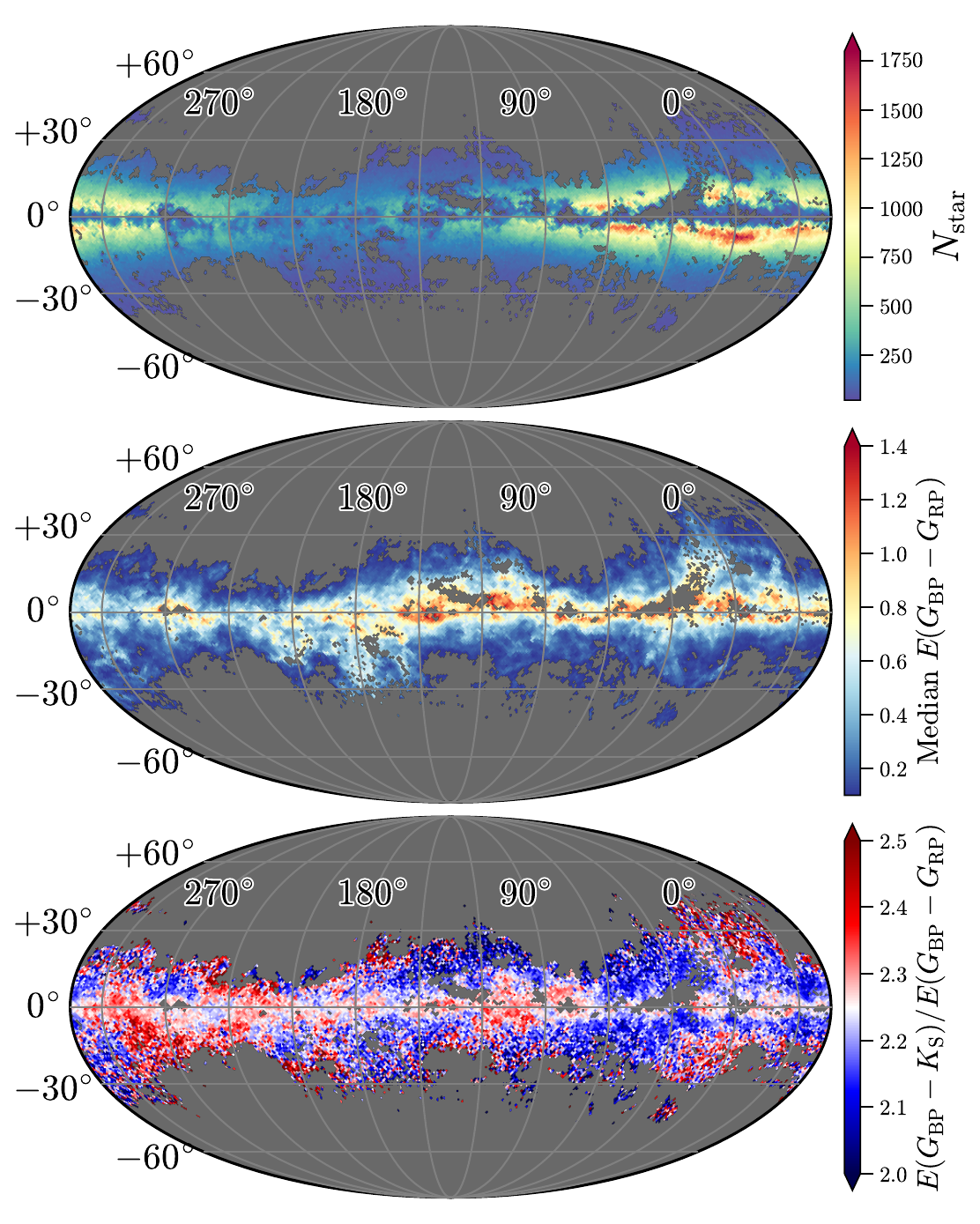}
  \caption{Two-dimensional Galactic distribution of the number of stars per pixel ($N_{\rm tar}$, top), the median $\EBPRP$ (middle),
  and the fit $\EBPK/\EBPRP$ (bottom) in 18\,236 HEALPixels with a pixel resolution of $55^{\prime}$ (see Sect. \ref{sect:cer}).}
  \label{fig:cer}
\end{figure*}

An extended application of the XGBoost reddening results is to derive the color excess ratio which can be used to characterize
the dust properties. As a preliminary study, we explored the 2D Galactic distribution of $\EBPK/\EBPRP$ based on the target sample 
built in Sect. \ref{subsect:ev-qc}. A linear fit was applied to $\EBPK$ and $\EBPRP$ in each spatial bin separated by HEALPix 
algorithm \citep{Gorski2005} with a resolution of $55^{\prime}$ ($N_{\rm side}\,{=}\,64$). Besides the slope ($\EBPK/\EBPRP$) and 
intercept from the linear fit, the number of target stars ($N_{\rm star}$), the median $\EBPRP$, and the Pearson correlation coefficient 
($r_p$) of the stars in each HEALPixel were calculated as well. Fit results in 18\,236 (37\%) HEALPixels satisfied the quality-control 
conditions: $N_{\rm star}\,{\geqslant}\,20$, $r_p\,{>}\,0.85$, median $\EBPRP\,{>}\,0.1$, and intercept within 0.1. The distribution 
of $N_{\rm star}$, median $\EBPRP$, and the fit $\EBPK/\EBPRP$ in these HEALPixels is shown in Fig. \ref{fig:cer}. Due to the controls 
made for the selections of the A23-RGB sample and the training-like sources, sources with high reddenings were almost dropped. Thus, 
nearby dense molecular regions, such as the famous Aquila Rift, were missed in the fit results. Based on the $N_{\rm tar}$ 
distribution, the sources in the target sample are most abundant with $|b|$ between $5^{\circ}$ and $10^{\circ}$ and toward the 
Galactic center. 
%---------------------------------------------------------------------------------------------------------------------------------

Despite the selection effect, the 2D distribution of the abundance of the dust grains is well described by the median $\EBPRP$
from diffuse molecular to translucent regions if we assume the cloud types proposed in \citet{SM2006} up to several kiloparsecs. 
$\EBPK/\EBPRP$ presents some large-scale variations with the Galactic longitude. For instance, $\EBPK/\EBPRP$ within $10^{\circ}$
to $50^{\circ}$ is remarkably lower than that between $90^{\circ}$ and $120^{\circ}$ or between $200^{\circ}$ and $240^{\circ}$ in
the middle plane. $\EBPK/\EBPRP$ is positively correlated with $\Rv$, the ratio of the total-to-selective extinction, but their
conversion is affected by stellar SED and dust reddening, especially in very broad passbands such as Gaia $G_{\rm BP}$ and $G_{\rm 
RP}$. Thus the accurate determination of $\Rv$ is beyond the scope of this work. Nevertheless, the $\EBPK/\EBPRP$ variation pattern 
shown in our results is very similar to the findings in \citet{Whittet1977}, \citet{Schlafly2016}, and \citet{ZhangRY2023} for $\Rv$. 
The structures are generally much larger than the scales of individual molecular clouds and their origin needs to be further explored 
with more bands.
%---------------------------------------------------------------------------------------------------------------------------------

\section{Summary} \label{sect:summary}

We developed an XGBoost model to predict the stellar intrinsic colors (ICs) with the atmospheric parameters $\teff$, $\logg$, and 
$\meta$, which is an improvement of the widely used blue-edge method that builds analytic relations between $\teff$ and ICs in large 
data sets. ICs in three bands were trained and tested in this work, that is $\BPRPO$, $\BPKO$, and $\JKO$, based on 1\,040\,446 
sources with low reddenings and high-quality stellar parameters and photometry selected from the Gaia--2MASS cross-matched catalog.
%---------------------------------------------------------------------------------------------------------------------------------

Stellar parameters used in the training and test sets were determined by the Gaia DR3 {\gspphot} module \citep{GSP-phot}. The accuracy
of {\gspphot} parameters were validated by comparing with APOGEE and LAMOST for 30\,130 and 243\,148 common sources, respectively.
The MedAD, RMSD, and AD90\% of the parameter differences significantly reduced compared to those for the whole {\gspphot} sample
containing highly reddened sources \citep{GSP-phot}. Moreover, the biases in {\gspphot} parameters, especially for $\meta$, do not 
have an apparent impact on the IC prediction as well. 
%---------------------------------------------------------------------------------------------------------------------------------

The generalization error of the XGBoost model, represented by RMSD in the test set, is 0.014\,mag for $\BPRPO$, 0.050\,mag for 
$\BPKO$, and 0.040\,mag for $\JKO$, which is on the same order of the magnitude as the photometric constraints for Gaia and 2MASS
bands. The model uncertainty also slightly decreases with $\teff$ and $\meta$ but increases with $\logg$ and $G$--magnitude.
%---------------------------------------------------------------------------------------------------------------------------------

We applied the XGBoost model to a target sample containing 5\,714\,528 stars selected from the RGB sample in \citetalias{Andrae2023}.
With derived ICs, the dust reddenings in corresponding bands, $\EBPRP$, $\EBPK$, and $\EJK$, were calculated. The high consistency
in the comparison of $\EJK$ between our results and literature values from \citetalias{ZhangXY2023}, \citetalias{Green2019}, and 
{\gspphot} further validates the accuracy of the XGBoost model. A significant overestimation of $\EJK$ is seen in \citetalias{Green2019} 
and {\gspphot} but not in \citetalias{ZhangXY2023} and our results. The XGBoost model slightly underestimated the IC of cool giants 
and consequently overestimated their reddenings because such stars were poorly covered by our training set. Additionally, we 
illustrated that the parameter distance ($d_P$) between the target sample and the training set can be used to select the training-like 
sources which can largely prevent the extrapolated prediction of the model.
%---------------------------------------------------------------------------------------------------------------------------------

As a representation of the extinction law and dust properties, the distribution of the color excess ratio, $\EBPK/\EBPRP$, was
preliminarily explored by linear fits done to $\EBPK$ and $\EBPRP$ of the target sample in HEALPixels with a spatial resolution of
$55^{\prime}$ ($N_{\rm side}\,{=}\,64$). Dense molecular regions were missed in the target sample due to the cuts for the A23-RGB 
sample and the quality control of $\EJK$. Nevertheless, the median $\EBPRP$ in selected HEALPixels well described the 2D dust 
distribution from diffuse molecular to translucent regions up to a few kiloparsecs. We further found a variation of $\EBPK/\EBPRP$
with the Galactic longitude in a similar pattern to the findings in \citet{Whittet1977}, \citet{Schlafly2016}, and \citet{ZhangRY2023} 
for $\Rv$.
%---------------------------------------------------------------------------------------------------------------------------------

Although there have been several Gaia-based studies on dust reddening, we focus on the training of ML models for IC predictions and
reddening calculations based on a vast sample of low-reddening stars from the Gaia homogeneous stellar parameterization. The trained
models can be applied to new spectro-photometric data, including designed observations in a small sample and the next-generation 
large spectroscopic surveys. The trained XGBoost model, the data sets used for training and testing, a catalog containing
a representative set of ICs as a function of stellar parameters, and a simple script showing how to use the model can be accessed 
via \url{https://doi.org/10.5281/zenodo.12787594}. The training of XGBoost models for other bands from optical to infrared will be 
completed in a follow-up work.
%---------------------------------------------------------------------------------------------------------------------------------

\begin{acknowledgments}
We thank the anonymous referee for very helpful suggestions and constructive comments.
This work is supported by the National Natural Science Foundation of China (NSFC) through the projects No. 12203099, 12373028, 12133002, and 12003046.
HZ acknowledges financial support of the China Postdoctoral Science Foundation (No. 2022M723373) and the Jiangsu Funding Program for Excellent Postdoctoral Talent.
SW acknowledges support from the Youth Innovation Promotion Association of the CAS (grant No. 2023065).
This work has made use of data from the European Space Agency (ESA) mission Gaia (\url{https://www.cosmos.esa.int/gaia}), 
processed by the Gaia Data Processing and Analysis Consortium (DPAC, \url{https://www.cosmos.esa.int/web/gaia/dpac/consortium}). 
Funding for the DPAC has been provided by national institutions, in particular, the institutions participating in the Gaia Multilateral Agreement.
\end{acknowledgments}
%---------------------------------------------------------------------------------------------------------------------------------

\vspace{5mm}
\facilities{Gaia, 2MASS, LAMOST, APOGEE}

\software{astropy \citep{astropy,astropy2},  
          dustmaps \citep{Green2018python},
          xgboost \citep{XGBoost},
          scikit-learn \citep{Pedregosa2011}}

\appendix

\section{Hyper-parameter optimization for XGBoost} \label{app:cv}

A 10-fold cross-validation, completed by the function `validation\_curve' in the Python package {\it scikit-learn} \citep{Pedregosa2011},
was used to optimize six hyper-parameters of the XGBoost model. For a given set of hyper-parameters, the training set will be
randomly separated into ten equal groups with nine for training and one as the validation set. After ten times, we use the mean 
and standard deviation ($std$) of the $R^2$ score of the training and validation set to select the best parameter. $R^2$ is 
defined as $1 - \frac{\sum(y-\hat{y})^2}{\sum(y-\bar{y})^2}$, where $y$ is the observed values, $\bar{y}$ is its mean, and $\hat{y}$ 
is the model prediction, the closer to 1 the better $R^2$. In this work, $R^2$ is the average score for the three colors. For each 
hyper-parameter, we test a sequence of values, keeping other parameters unchanged, to find the best value. If the hyper-parameter gets 
a new value compared to the last round, the optimization process will be done again for all the others until the best combination of 
the hyper-parameters is achieved. 
%---------------------------------------------------------------------------------------------------------------------------------

The final combination of the hyper-parameters is listed in Sect. \ref{subsect:model} and their validation curves are presented 
in Fig. \ref{fig:cv-curve}. The validation curves show the mean and {\it std} of $R^2$ (blue for the training set and orange for 
the validation set) for a set of values around the best one for each hyper-parameter. For max\_depth, both training and CV $R^2$ 
increase until six and then the training $R^2$ keeps growing but the CV $R^2$ decreases with larger max\_depth, indicating an
overfitting of the model. Therefore, $\rm max\_depth\,{=}\,6$ was chosen. For $\rm n\_estimators\,{>}\,150$, both training and CV
$R^2$ are no longer sensitive to n\_estimators, similar to that for reg\_lambda and max\_leaves. The learning\_rate is 0.1 when
both training and CV $R^2$ reach their maximum. On the other hand, both training and CV $R^2$ decrease with gamma. We selected
$\rm gamma\,{=}\,0.02$ so that the training and CV $R^2$ are relatively high and not too far away from each other. More sophisticated
model selection could be applied but it will take a much longer time and the model performance would not improve much.
%---------------------------------------------------------------------------------------------------------------------------------

\section{Python script to use the trained XGBoost model} \label{app:script}

\begin{lstlisting}[language=Python]
import pickle
from astropy.io import fits

def use_xgb(teff, logg, mh):
    '''
    -------------------------------------------
    Usage:   
                - Use the trained XGBoost model 
                  to predict stellar intrinsic
                  colors with stellar atmospheric 
                  parameters.
    Input:
         teff:  - effective temperature,
                  (n,) numpy array.
         logg:  - logarithmic gravit,
                  (n,) numpy array.
         mh:    - metallicity [M/H],
                  (n,) numpy array.
    
    Output:
         ypred: - predicted intrinsic colors,
                  (n,3) numpy array, 
                  Column 1: BP-RP
                  Column 2: BP-Ks
                  Column 3: J-Ks
        
    Author:             He Zhao
    Email:              he.zhao@oca.eu
    Finished:           2024.06.25
    Update Log:
    -------------------------------------------
    '''
    filename = 'user_path/xgb_model.pkl'
    xgb = pickle.load(open(filename,'rb'))
    X   = np.vstack((teff,logg,mh)).T
    ypred = xgb.predict(X)
    return ypred
\end{lstlisting}
%---------------------------------------------------------------------------------------------------------------------------------

\section{Performance of XGBoost on test set} \label{app:test}

As supplementary materials for the discussions in Sect. \ref{subsect:test_set}, Fig. \ref{fig:test-set-sub} presents the variation
of the statistics (MedAD, RMSD, AD90\%) of the color difference in the test set as a function of the stellar parameters and 
$G$--magnitude. The bin step is 200\,K for $\teff$, 0.25 for $\logg$, 0.25\,dex for $\meta$, and 0.5\,mag for $G$. Extreme values
and large fluctuations mostly occur in bins with insufficient sources. Figure \ref{fig:test-set-hot} shows the color differences
as a function of stellar parameters and $G$ for hot stars between 8000 and 20\,000\,K.
%---------------------------------------------------------------------------------------------------------------------------------

\bibliography{sample631}{}
\bibliographystyle{aasjournal}

\begin{deluxetable}{ccrccc}
\tablecaption{Intrinsic colors at typical stellar parameter values predicted by the XGBoost model. The selection of the parameter 
combinations is stated in Sect. \ref{subsect:model}. \label{tab:IC}}
\tablehead{
\colhead{$\teff$} & \colhead{$\logg$} & \multicolumn{1}{c}{$\meta$} & \colhead{$\BPRPO$} & \colhead{$\BPKO$} & \colhead{$\JKO$} \\
\colhead{(K)}     &                   & \colhead{(dex)}   & \colhead{(mag)}    & \colhead{(mag)}   & \colhead{(mag)}
}
\startdata
4200 &	1 &  --1.0 & 1.46 & 3.33 & 0.80 \\
4400 &	1 &    0.4 & 1.46 & 3.31 & 0.81 \\
4600 &	1 &    0.4 & 1.36 & 3.04 & 0.73 \\
5000 &	2 &  --2.0 & 1.02 & 2.31 & 0.55 \\
4400 &	2 &  --1.0 & 1.30 & 2.94 & 0.71 \\
4600 &	2 &  --1.0 & 1.20 & 2.69 & 0.64 \\
4800 &	2 &  --1.0 & 1.11 & 2.52 & 0.60 \\
5000 &	2 &  --1.0 & 1.03 & 2.33 & 0.55 \\
5200 &	2 &  --1.0 & 0.96 & 2.17 & 0.51 \\
4400 &	2 &	   0.0 & 1.40 & 3.13 & 0.75 \\
4600 &	2 &	   0.0 & 1.28 & 2.87 & 0.67 \\
4800 &	2 &	   0.0 & 1.18 & 2.65 & 0.61 \\
5400 &	3 &	 --2.0 & 0.87 & 1.93 & 0.44 \\
5600 &	3 &	 --2.0 & 0.80 & 1.78 & 0.40 \\
5000 &	3 &	 --1.0 & 1.03 & 2.27 & 0.54 \\
5200 &	3 &	 --1.0 & 0.96 & 2.14 & 0.50 \\
5400 &	3 &	 --1.0 & 0.90 & 1.96 & 0.46 \\
5600 &	3 &	 --1.0 & 0.83 & 1.81 & 0.42 \\
5800 &	3 &	 --1.0 & 0.77 & 1.62 & 0.37 \\
6200 &	3 &	 --1.0 & 0.66 & 1.39 & 0.31 \\
6400 &	3 &	 --1.0 & 0.61 & 1.28 & 0.27 \\
7400 &	3 &	 --1.0 & 0.33 & 0.71 & 0.15 \\
4600 &	3 &	   0.0 & 1.29 & 2.86 & 0.66 \\
4800 &	3 &	   0.0 & 1.18 & 2.58 & 0.60 \\
5000 &	3 &	   0.0 & 1.10 & 2.39 & 0.55 \\
5200 &	3 &	   0.0 & 1.03 & 2.26 & 0.51 \\
5400 &	3 &	   0.0 & 0.97 & 2.08 & 0.47 \\
6200 &	3 &	   0.0 & 0.70 & 1.47 & 0.31 \\
6000 &	4 &	 --2.0 & 0.72 & 1.65 & 0.35 \\
6200 &	4 &	 --2.0 & 0.66 & 1.52 & 0.31 \\
6400 &	4 &  --2.0 & 0.62 & 1.40 & 0.29 \\
7000 &	4 &	 --2.0 & 0.47 &	1.07 & 0.20 \\
7200 &	4 &	 --2.0 & 0.42 &	0.96 & 0.18 \\
7400 &	4 &	 --2.0 & 0.37 &	0.86 & 0.16 \\
7600 &	4 &	 --2.0 & 0.32 &	0.80 & 0.14 \\
\enddata
\end{deluxetable}\addtocounter{table}{-1}
%---------------------------------------------------------------------------------------------------------------------------------

\begin{deluxetable}{ccrccc}
\tablecaption{--continued.}
\tablehead{
\colhead{$\teff$} & \colhead{$\logg$} & \multicolumn{1}{c}{$\meta$} & \colhead{$\BPRPO$} & \colhead{$\BPKO$} & \colhead{$\JKO$} \\
\colhead{(K)}     &                   & \colhead{(dex)}   & \colhead{(mag)}    & \colhead{(mag)}   & \colhead{(mag)}
}
\startdata
3200 & 4 & --1.0 & 2.46 & 4.84 & 0.84 \\
3400 & 4 & --1.0 & 2.12 & 4.24 & 0.83 \\
3600 & 4 & --1.0 & 1.89 & 3.94 & 0.84 \\
5800 & 4 & --1.0 & 0.78 & 1.71 & 0.38 \\
6000 & 4 & --1.0 & 0.72 & 1.57 & 0.34 \\
6200 & 4 & --1.0 & 0.66 & 1.45 & 0.30 \\
6400 & 4 & --1.0 & 0.62 & 1.34 & 0.28 \\
6600 & 4 & --1.0 & 0.57 & 1.22 & 0.25 \\
6800 & 4 & --1.0 & 0.52 & 1.11 & 0.22 \\
7000 & 4 & --1.0 & 0.47 & 1.00 & 0.19 \\
7200 & 4 & --1.0 & 0.42 & 0.89 & 0.17 \\
7400 & 4 & --1.0 & 0.37 & 0.79 & 0.15 \\
7600 & 4 & --1.0 & 0.32 & 0.73 & 0.13 \\
7800 & 4 & --1.0 & 0.29 & 0.62 & 0.10 \\
8000 & 4 & --1.0 & 0.21 & 0.47 & 0.08 \\
9000 & 4 & --1.0 & 0.08 & 0.18 & 0.02 \\
10000\,\,\, &	4 &	--1.0 &	--0.03\,\,\, & --0.02\,\,\, & --0.01\,\,\, \\
11000\,\,\, &	4 &	--1.0 &	--0.08\,\,\, & --0.14\,\,\, & --0.04\,\,\, \\
12000\,\,\, &	4 &	--1.0 &	--0.11\,\,\, & --0.24\,\,\, & --0.07\,\,\, \\
4200 & 4 & 0.0 & 1.54 &	3.33 & 0.76 \\
4400 & 4 & 0.0 & 1.38 &	2.98 & 0.70 \\
4800 & 4 & 0.0 & 1.15 &	2.50 & 0.57 \\
5000 & 4 & 0.0 & 1.07 &	2.32 & 0.52 \\
5200 & 4 & 0.0 & 0.99 &	2.12 & 0.47 \\
5400 & 4 & 0.0 & 0.93 &	1.97 & 0.43 \\
5600 & 4 & 0.0 & 0.87 &	1.83 & 0.39 \\
5800 & 4 & 0.0 & 0.78 &	1.63 & 0.34 \\
6000 & 4 & 0.0 & 0.72 &	1.49 & 0.30 \\
6200 & 4 & 0.0 & 0.66 &	1.35 & 0.27 \\
6400 & 4 & 0.0 & 0.61 &	1.25 & 0.24 \\
6600 & 4 & 0.0 & 0.56 &	1.13 & 0.22 \\
6800 & 4 & 0.0 & 0.52 &	1.04 & 0.19 \\
7000 & 4 & 0.0 & 0.46 &	0.92 & 0.17 \\
7200 & 4 & 0.0 & 0.41 &	0.81 & 0.14 \\
7400 & 4 & 0.0 & 0.36 &	0.70 & 0.12 \\
7600 & 4 & 0.0 & 0.32 &	0.64 & 0.10 \\
\enddata
\end{deluxetable}\addtocounter{table}{-1}
%---------------------------------------------------------------------------------------------------------------------------------

\begin{deluxetable}{ccrccc}
\tablecaption{--continued.}
\tablehead{
\colhead{$\teff$} & \colhead{$\logg$} & \multicolumn{1}{c}{$\meta$} & \colhead{$\BPRPO$} & \colhead{$\BPKO$} & \colhead{$\JKO$} \\
\colhead{(K)}     &                   & \colhead{(dex)}   & \colhead{(mag)}    & \colhead{(mag)}   & \colhead{(mag)}
}
\startdata
7800 & 4 & 0.0 & 0.28 &	0.52 & 0.07 \\
9000 & 4 & 0.0 & 0.08 &	0.09 & 0.00 \\
10000\,\,\, & 4 & 0.0 & --0.04\,\,\, & --0.11\,\,\, & --0.03\,\,\, \\
11000\,\,\, & 4 & 0.0 & --0.09\,\,\, & --0.23\,\,\, & --0.06\,\,\, \\
15000\,\,\, & 4 & 0.0 & --0.21\,\,\, & --0.48\,\,\, & --0.12\,\,\, \\
18000\,\,\, & 4 & 0.0 & --0.27\,\,\, & --0.59\,\,\, & --0.13\,\,\, \\
5000 & 4 & 0.4 & 1.09 &	2.32 & 0.52 \\
5200 & 4 & 0.4 & 1.01 &	2.12 & 0.47 \\
5400 & 4 & 0.4 & 0.94 &	1.96 & 0.42 \\
5600 & 4 & 0.4 & 0.87 &	1.81 & 0.38 \\
5800 & 4 & 0.4 & 0.79 &	1.64 & 0.33 \\
6000 & 4 & 0.4 & 0.73 &	1.51 & 0.30 \\
6200 & 4 & 0.4 & 0.67 &	1.39 & 0.26 \\
6600 & 4 & 0.4 & 0.57 &	1.17 & 0.21 \\
6800 & 4 & 0.4 & 0.52 &	1.06 & 0.19 \\
7000 & 4 & 0.4 & 0.47 &	0.94 & 0.16 \\
7200 & 4 & 0.4 & 0.42 &	0.81 & 0.14 \\
7800 & 4 & 0.4 & 0.29 &	0.53 & 0.06 \\
11000\,\,\, & 4 & 0.4 & --0.09\,\,\, & --0.22\,\,\, & --0.06\,\,\, \\
12000\,\,\, & 4 & 0.4 & --0.11\,\,\, & --0.32\,\,\, & --0.09\,\,\, \\
13000\,\,\, & 4 & 0.4 & --0.16\,\,\, & --0.42\,\,\, & --0.09\,\,\, \\
14000\,\,\, & 4 & 0.4 & --0.19\,\,\, & --0.47\,\,\, & --0.12\,\,\, \\
20000\,\,\, & 4 & 0.4 & --0.26\,\,\, & --0.58\,\,\, & --0.13\,\,\, \\
3200 & 5 & --1.0 & 2.60 & 4.68 & 0.73 \\
3800 & 5 & --1.0 & 1.74 & 3.49 & 0.73 \\
4000 & 5 & --1.0 & 1.59 & 3.30 & 0.67 \\
3000 & 5 &   0.0 & 3.31 & 6.16 & 0.89 \\
3200 & 5 &   0.0 & 2.83 & 5.47 & 0.85 \\
3400 & 5 &   0.0 & 2.49 & 5.04 & 0.85 \\
3000 & 5 &   0.4 & 3.36 & 6.34 & 0.90 \\
\enddata
\end{deluxetable}
%---------------------------------------------------------------------------------------------------------------------------------

%% The "ht!" tells LaTeX to put the figure "here" first, at the "top" next
%% and to override the normal way of calculating a float position
%---------------------------------------------------------------------------------------------------------------------------------

%---------------------------------------------------------------------------------------------------------------------------------

%---------------------------------------------------------------------------------------------------------------------------------

%---------------------------------------------------------------------------------------------------------------------------------

%---------------------------------------------------------------------------------------------------------------------------------

%---------------------------------------------------------------------------------------------------------------------------------

%---------------------------------------------------------------------------------------------------------------------------------

%---------------------------------------------------------------------------------------------------------------------------------

%---------------------------------------------------------------------------------------------------------------------------------
% Figures in appendix
%---------------------------------------------------------------------------------------------------------------------------------

\begin{figure*}[ht!]
  \centering
  \includegraphics[width=16cm]{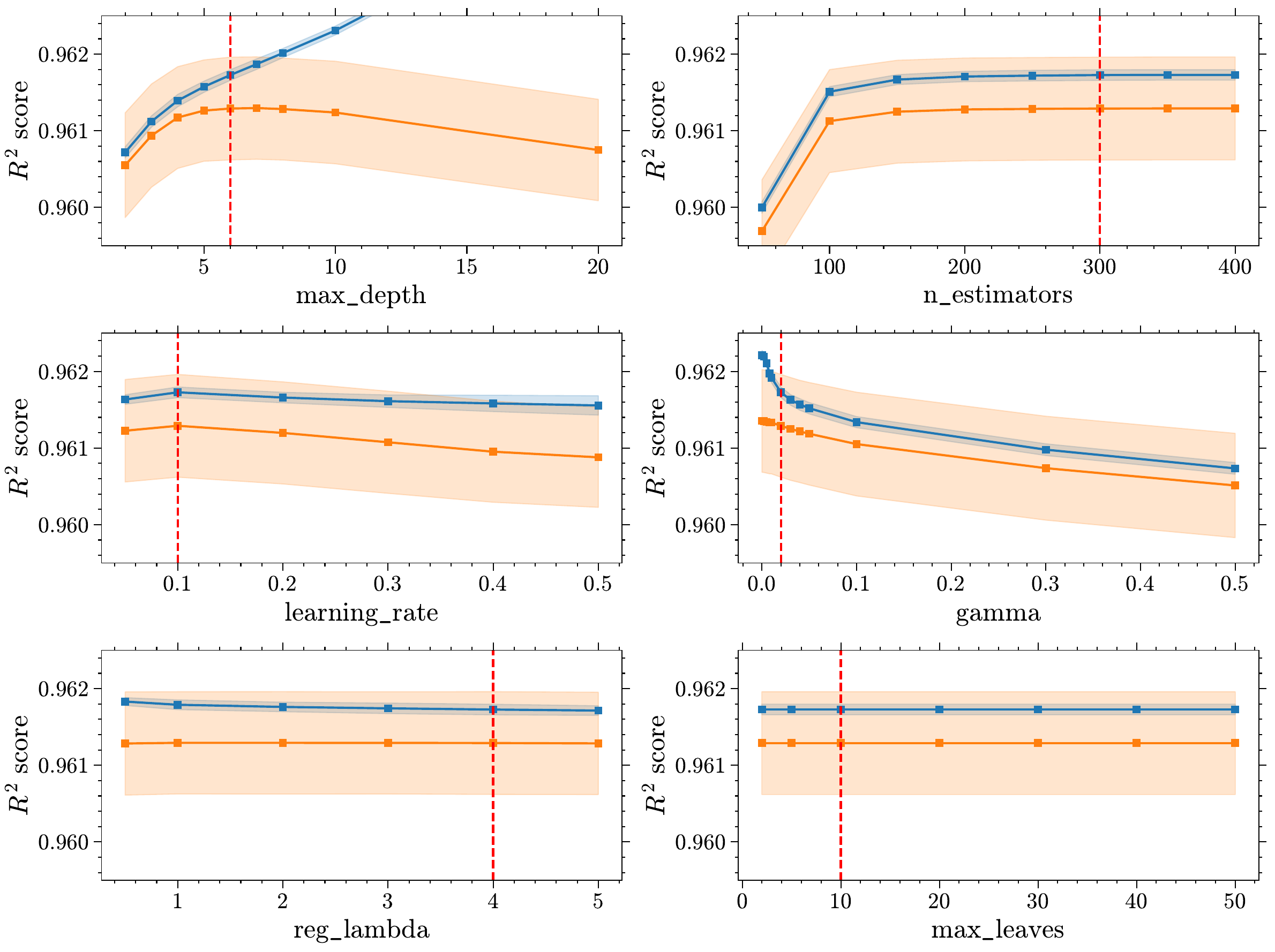}
  \caption{Validation curves of the optimized hyper-parameters for the XGBoost model (see Sect. \ref{subsect:model}). The blue and 
  orange lines show the mean $R^2$ score in the 10-fold cross-validation for the training and validation sets, respectively, with 
  a span of their standard deviation. The squares indicate the mean $R^2$ score at specific values of the hyper-parameters, and the 
  dashed red lines mark the finally chosen hyper-parameters.}
  \label{fig:cv-curve}
\end{figure*}
%---------------------------------------------------------------------------------------------------------------------------------

\begin{figure*}[ht!]
  \centering
  \includegraphics[width=16cm]{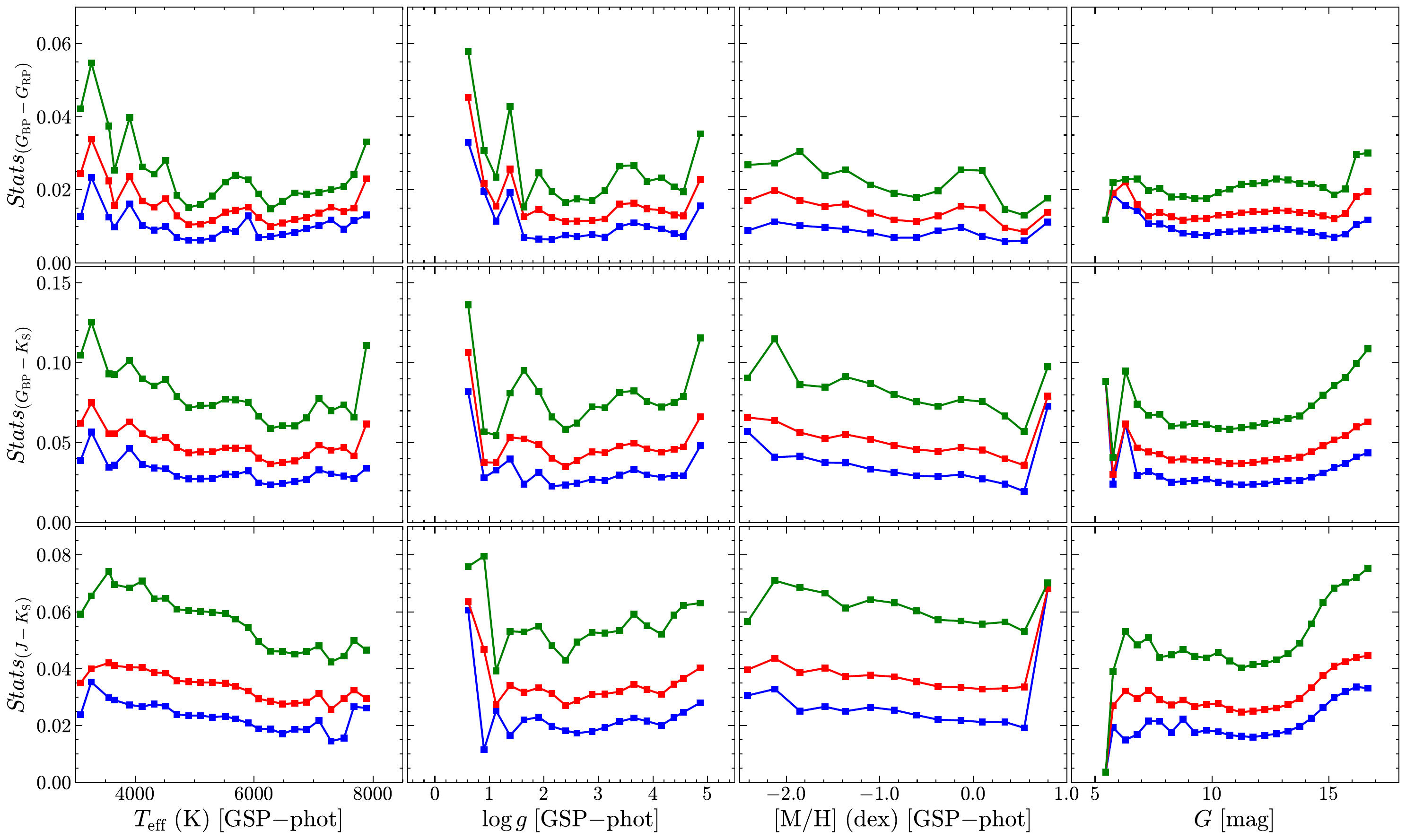}
  \caption{Statistics of the absolute color differences between observed colors and the XGBoost predictions as a function for $\teff$, 
  $\logg$, $\meta$, and $G$--magnitude for the test set. Rows from top to bottom show the differences for $\BPRP$, $\BPK$, and $\JK$.
  For colored squares and lines in each panel: blue is the median absolute difference (MedAD), red is the root-mean-square difference 
  (RMSD), and green is the absolute difference not exceeded by 90\% of sources (AD90\%). These statistics were calculated in different
  bins with a step of 200\,K for $\teff$, 0.25 for $\logg$, 0.25\,dex for $\meta$, and 0.5\,mag for $G$, respectively.}
  \label{fig:test-set-sub}
\end{figure*}
%---------------------------------------------------------------------------------------------------------------------------------

\begin{figure*}[ht!]
  \centering
  \includegraphics[width=16cm]{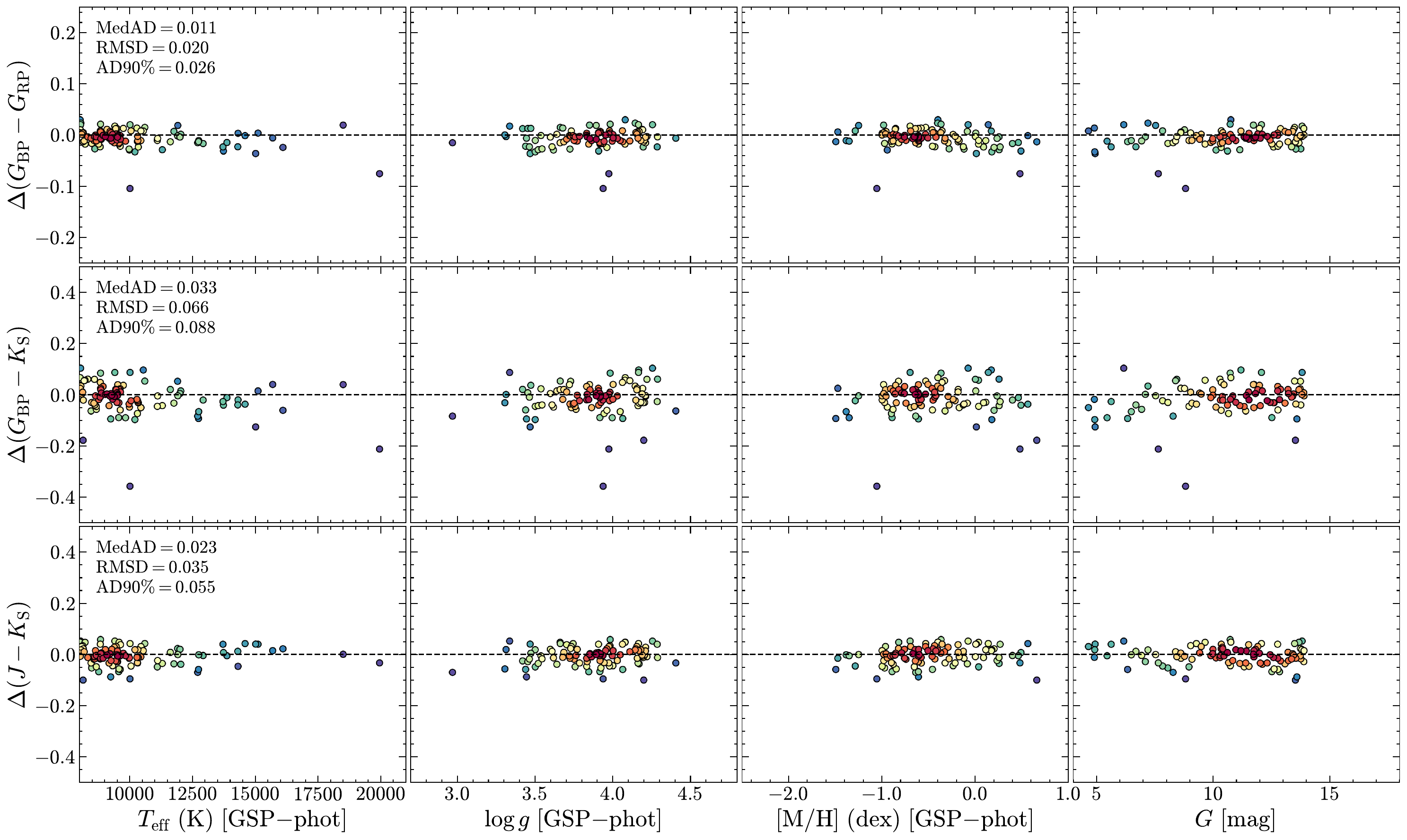}
  \caption{The same as Fig. \ref{fig:test-set} but for stars hotter than 8000\,K. The points are color-coded by their number
  density calculated by a Gaussian kernel density estimation.}
  \label{fig:test-set-hot}
\end{figure*}
%---------------------------------------------------------------------------------------------------------------------------------

\clearpage
\end{CJK*}
\end{document}